\documentclass[aj]{emulateapj}
\usepackage{apjfonts}

\slugcomment{To be published in the Astronomical Journal}
\received{Aug 24, 2004}
\revised{Oct 13, Nov 12, 2004}
\accepted{Nov 18, 2004}

\def\hst{{\sl HST}}
\def\iras{{\sl IRAS}}

\def\msun{M$_{\odot}$}
\def\ipol{$I_{\rm p}$}
\def\kms{km s$^{-1}$}
\makeatletter
\def\ale{\mathrel{\mathpalette\gl@align<}}
\def\age{\mathrel{\mathpalette\gl@align>}}
\def\gl@align#1#2{\lower.6ex\vbox{\baselineskip\z@skip\lineskip\z@
\ialign{$\m@th#1\hfil##\hfil$\crcr#2\crcr\sim\crcr}}}
\makeatletter

\shorttitle{HST/NICMOS Imaging Polarimetry of PPNs} 
\shortauthors{Ueta, Murakawa, \& Meixner}

\begin{document}
 
\title{HST/NICMOS Imaging Polarimetry 
of Proto-Planetary Nebulae:\\
Probing of the Dust Shell Structure via Polarized Light} 

\author{%
Toshiya Ueta\altaffilmark{1,2,3},
Koji Murakawa\altaffilmark{4,5},
Margaret Meixner\altaffilmark{6}}

\altaffiltext{1}{Royal Observatory of Belgium,
Ringlaan, 3, 
B-1180, Brussels, Belgium}

\altaffiltext{2}{Current Address:
NASA Ames Research Center/SOFIA,
Mail Stop 211-3,
Moffett Field, CA 94035, USA;
tueta@mail.sofia.usra.edu}

\altaffiltext{3}{NRC Research Associate}

\altaffiltext{4}{Subaru Telescope,
National Astronomical Observatory of Japan,
650 North A'ohoku Place 
Hilo, HI 96720, USA;
murakawa@subaru.naoj.org}

\altaffiltext{5}{Current Address:
ASTRON, 
P.O.\ Box 2, 7990 AA, Dwingeloo
The Netherlands}

\altaffiltext{6}{Space Telescope Science Institute, 
3700 San Martin Drive, Baltimore, MD 21218, USA;
meixner@stsci.edu}

\begin{abstract}
Using NICMOS on {\hst}, we have performed imaging 
polarimetry of proto-planetary nebulae.
Our objective is to study the structure of 
{\sl optically thin} circumstellar shells of 
post-asymptotic giant branch stars by separating
dust-scattered, linearly polarized star light 
from unpolarized direct star light.
This unique technique allows us to probe faint 
reflection nebulae around the bright central 
star, which can be buried under the 
point-spread-function of the central star
in conventional imaging.
Our observations and archival search have yielded 
polarimetric images for five sources: 
{\iras} 07134$+$1005 (HD 56126), 
{\iras} 06530$-$0213, 
{\iras} 04296$+$3429, 
{\iras} (Z)02229$+$6208, and 
{\iras} 16594$-$4656.
These images have revealed the circumstellar dust 
distribution in an unprecedented detail via polarized 
intensity maps, providing a basis to understand 
the 3-D structure of these dust shells.   
We have observationally confirmed 
the presence of the inner cavity caused by the cessation 
of AGB mass loss and the internal shell structures which 
is strongly tied to the progenitor star's mass loss history
on the AGB.
We have also found that equatorial enhancement in 
these circumstellar shells comes with various degrees 
of contrast, suggesting a range of optical depths in
these optically thin shells.
Our data support the interpretation that the dichotomy 
of PPN morphologies is due primarily to differences in 
optical depth and secondary to the inclination effect.
The polarization maps reveal a range of inclination 
angles for these optically thin reflection nebulae, 
dispelling the notion that elliptical nebulae are 
pole-on bipolar nebulae.
\end{abstract}


\section{Introduction}

The proto-planetary nebula (PPN) phase is a relatively 
short ($\sim 10^{3}$ years) stage of stellar evolution 
for low to intermediate initial mass ($\sim 0.8 - 8$ \msun) 
stars between the asymptotic giant branch (AGB) and 
planetary nebula (PN) phases (e.g., \citealt{kwok93,vanwinckel03}).
During the PPN phase the post-AGB central star increases 
its surface temperature from a few to a few tens of $10^3$ 
K, while the circumstellar dust shell - created by the 
AGB mass loss and physically detached from the central
star at the end of the AGB phase - simply coasts away 
from the central star.
Therefore, PPNs are important stellar objects in which 
to investigate the nature of dusty mass loss during the 
AGB phase, because the most pristine history of AGB mass 
loss is imprinted and preserved in their density distribution.

The AGB mass loss history can provide crucial 
clues for the shell structure formation.
While the circumstellar shells of AGB stars initially assume 
spherically symmetric shape when they are formed, they seem
to develop largely axisymmetric structure by the time the 
AGB mass loss is terminated (e.g., \citealt{balick02} for
a review).
Meanwhile, the AGB mass loss history can also yield
information about the internal evolution of the central 
star.
AGB mass loss may show temporal variations due to the 
alternative burning of hydrogen and helium in two distinct 
layers via the mechanism called thermal pulsation \citep{iben81}, 
while a sudden enhancement of mass loss near the end of the 
AGB phase - the so-called superwind \citep{renzini81,iben83} - 
may remove almost the entire surface layer from the star 
terminating the AGB evolution.
Thus, the AGB mass loss history is strongly linked to both 
the internal and external evolution of the central star,
and the effects of AGB mass loss can manifest themselves 
in the density distribution in the PPN dust shells.

The density distribution in PPNs can be observationally 
studied in two ways: either directly via thermal dust 
emission arising from the shells or indirectly via 
scattered light through the dusty shells.
PPN imaging surveys were conducted using both methods,
and the combined results suggested that PPNs were 
intrinsically axisymmetric due to equatorially-enhanced 
mass loss near the end of the AGB phase, which probably 
coincides with the superwind phase \citep{meixner99,ueta00}.
These surveys also found a morphological dichotomy among 
PPNs, showing one-to-one correspondence between the mid-IR 
and optical morphologies: an optical bipolar reflection 
nebula is always found with a mid-IR emission nebula of 
a single core surrounded by an elliptical low-emission halo 
(DUPLEX-core/elliptical PPNs), while an optical elliptical 
nebula is usually associated with a mid-IR emission nebula 
harboring two emission peaks as evidence for limb-brightened 
edge-on dust torus (SOLE-toroidal PPNs).
Subsequent radiative transfer calculations showed that the 
optical depth of the shell would play a more important role 
in determining the morphology of the shell than the 
inclination effect \citep{meixner02,ueta03}.

In both morphological cases, the source of axisymmetric 
structure is an equatorial density enhancement present 
in the innermost regions of the shell.
Hence, the AGB mass loss history represented by this part 
of the shell holds the critical piece of information to 
enhance our understanding of mass loss processes.
Taking advantage of their optically thin nature, 
SOLE-toroidal PPNs have been studied by high-resolution 
imaging of thermal dust emission at mid-IR in order to 
reveal the density distribution in these shells 
(\citealt{dayal98,ueta01,kwok02,gledhill03}).
However, such studies have been difficult to perform at 
sufficient spatial resolution because of the 
diffraction-limited 
nature of the mid-IR imaging and the intrinsically 
compact nature of PPNs.

In this paper, we use near-IR imaging polarimetry to 
investigate the structure of five SOLE-toroidal PPN: 
{\iras} 07134$+$1005 (HD 56126), 
{\iras} 06530$-$0213, 
{\iras} 04296$+$3429, 
{\iras} (Z)02229$+$6208, and 
{\iras} 16594$-$4656.
Below, we describe the technique of near-IR imaging 
polarimetry and motivate our use of it in this study 
(\S 2) and outline our observational procedure and 
data processing steps (\S 3).  
In \S 4, we present the results, and discuss the 
individual sources in \S 5.  
We then discuss some of the implications of the 
results (\S 6) and summarize the conclusions (\S 7).

\section{Imaging Polarimetry of PPNs}

\citet{gledhill01} used near-IR imaging polarimetry 
as an alternative technique to directly capture the 
structure of the PPN dust shells.
The imaging polarimetry can separate dust-scattered 
light (as the linearly polarized component) from 
direct star light (as the unpolarized component).
Hence, one can easily detect faint, dust-scattered 
light from PPNs that is otherwise buried beneath the 
dominant point-spread-function (PSF) of the central 
star: the PPN structure can thus be revealed as in 
the mid-IR imaging, but at one order better 
resolution in the near-IR.
For example, the shell structure of HD 235858 ({\iras} 
22272+5435), which was not fully resolved at mid-IR 
until observed with the 10 m Keck telescope 
\citep{ueta01}, was resolved by near-IR imaging 
polarimetry with the 4 m UKIRT \citep{gledhill01}.
Therefore, we performed imaging polarimetry with 
the Hubble Space Telescope ({\hst}\/) to investigate 
the structure of very compact PPNs by exploiting the 
unique ability of this novel observing method and 
the high resolution capabilities of the telescope.

Polarimetric characteristics of astronomical objects 
can be obtained by means of the Stokes parameters, 
($I$, $Q$, $U$\/), which are computed from the 
measurements of the beam intensity passing through 
linear polarizing elements.
The near-IR camera and multiple object spectrometer 
(NICMOS; \citealt{nicmosihb}) on {\hst} is equipped 
with three polarizers, and the Stokes parameters can 
be derived by a matrix conversion method that is
elucidated by \citet{sparks99} and \citet{hines00}.
Using the Stokes $I$ (the total intensity) along 
with the Stokes $Q$ and $U$, we can then express 
the intensity of linearly polarized light, {\ipol},
the polarization strength, $P$, and the polarization 
PA, $\theta$.

However, one must exercise caution when computing 
$P$ because $I$ can be affected by the (unpolarized) 
PSF of the central source (e.g., \citealt{gledhill01b}).
The PSF of the central star can
(1) accompany enormous spider structures that spatially 
affect the detectability of the circumstellar shell 
and
(2) induce very large $I$ with respect to potentially 
small {\ipol},
especially when the optical depth of the surrounding 
matter is low.
In such cases, the PSF contribution, 
$I_{\rm psf}$, has to be removed from $I$ to obtain 
the {\sl PSF-corrected} polarization strength,
$P_{\rm corr}$;
\begin{equation}\label{correctedP}
 P_{\rm corr} = \frac{I_{\rm p}}{I - I_{\rm psf}}.          
\end{equation}

In order for the PSF correction to be effective,
PSF reference observations must be carefully
designed.
The PSF observations should achieve a similar 
S/N to the source observations, yielding
spatially and quantitatively equivalent PSF 
structures.
The removal of the PSF effect is difficult even 
with good PSF reference data, since there is no 
way to know {\sl a priori} how much matter
exists along the line of sight in front of the 
central source.
The scaling of the PSF reference data with respect 
to the target data would always be a problem, and 
hence, $P$ as a ratio of {\ipol} to $I$ (irrespective 
of the PSF correction) should be considered to roughly
define the {\sl lower} limit.
In the following, we refer to $P_{\rm corr}$ as $P$.
On the contrary, {\ipol} and $\theta$ can be 
extracted even from the PSF-affected data
because these quantities depend only on 
$Q$ and $U$ which are free from the unpolarized 
component by definition.
Therefore, we will make use of the {\ipol} maps
in addition to the $P$ maps to investigate the
shell structure in PPNs.

\section{Observations and Data Reduction}

\subsection{Observations}

We obtained high-resolution near-IR polarimetric data 
from four PPNs using NICMOS on-board {\hst} through a general 
observer (GO) program 9377 (PI: T.\ Ueta) during Cycle 11, 
after the NICMOS cooling system (NCS) was installed during 
the servicing mission 3B (SM3B). 
We selected the target sources ({\iras} 07134$+$1005, {\iras} 
06530$-$0213, {\iras} 04296$+$3429, and {\iras}
[Z]02229$+$6208\footnote{%
Hereafter, we refer to each object by the RA part of the {\iras}
designation.
The prefix ``Z'' for {\iras} 02229 is given by \citet{hrivnak99}
due to the fact that the source was found in the Faint Source 
Reject File in the {\iras} Faint Source Survey.}) 
based on their morphological classification in a previous WFPC2 
survey \citep{ueta00}: they are all SOLE-toroidal PPNs.
We observed our targets with Camera 1 (NIC1), which provides 
$11\arcsec \times 11\arcsec$ field of view at $0\farcs043$ 
pixel$^{-1}$ scale, and short wavelength polarizers (POL0S, 
POL120S, and POL240S, covering 0.8 to 1.3 \micron). 
We used the MULTIACCUM (non-destructive readout) mode to 
recover information in saturated pixels from unsaturated 
readouts. 

To remove the effects of bad pixels and improve the spatial 
sampling of the PSF, we employed the dithering technique 
and observed each source at five or four dithering positions 
(i.e., four-point spiral pattern with or without the pattern 
center) with offsets by non-integer pixels.
At each dither position we took images using all three polarizers, 
allocating equal amount of exposure time to each polarizer.
The observations were designed this way to make the visits 
as efficient as possible.
However, taking multiple exposures at each dither position 
can cause a problem particularly when a wide dynamic range 
has to be covered, since image persistence due to very 
bright parts of the source (the central star, in our case) 
may leave photon persistence that can affect the subsequent 
exposures (e.g., \citealt{nicmoshandbook}).
To reduce the effect of persistence, we took two 5.158 sec 
BLANK exposures after each exposure with a polarizer.

\begin{deluxetable*}{lcccccccrc}[h]
\tablecolumns{10} 
\tablewidth{0pt} 
\tablecaption{\label{log}%
Log of {\hst}/NICMOS Observations} 
\tablehead{%
\colhead{} & 
\colhead{} &
\colhead{} &
\colhead{} &
\colhead{} & 
\multicolumn{2}{c}{DITHER} &
\colhead{EXPTIME\tablenotemark{b}} &
\colhead{ORIENTAT\tablenotemark{c}} &
\colhead{}\\
\cline{6-7}
\colhead{Source} &
\colhead{Date} &
\colhead{FILTER\tablenotemark{a}} &
\colhead{SAMP\_SEQ} &
\colhead{NSAMP} &
\colhead{PATTERN} &
\colhead{NPTS} &
\colhead{(sec)} &
\colhead{(deg)} &
\colhead{Ref.}} 

\startdata 
\cutinhead{New Data}

IRAS 07134$+$1005 & 
2003 Mar 29 & 
POL-S &
STEP8 & 
19 &
SPIRAL &
5 &
\phn519.5079\phn & 
$-$129.432\phn &
\\

IRAS 06530$-$0213 & 
2003 Mar 29 & 
POL-S &
STEP8 & 
19 &
SPIRAL &
5 &
\phn519.5079\phn & 
$-$131.793\phn &
\\

IRAS 04296$+$3429 & 
2003 Mar 28 & 
POL-S &
STEP8 & 
19 &
SPIRAL &
5 &
\phn519.5079\phn & 
$-$149.313\phn & 
\\

IRAS (Z)02229$+$6208\tablenotemark{d} & 
2003 Mar 28 & 
POL-S &
STEP8 & 
16 &
SPIRAL &
4 &
\phn319.6832\phn & 
169.035\phn &
\\

HD 12088\tablenotemark{e} & 
2003 Mar 28 & 
POL-S &
STEP1 & 
19 &
SPIRAL &
5 &
\phn\phn79.7868\phn &
162.51\phn\phn &
\\

\cutinhead{Archived Data}

IRAS 16594$-$4656 & 
1998 May \phn2 & 
POL-L &
STEP32 & 
17 &
SPIRAL &
3 &
\phn863.8767\phn  &
78.9531 &
2 \\



BD $+$32$^{\circ}$3739\tablenotemark{f} & 
1997 Sep \phn1 & 
POL-L &
STEP1 & 
\phn9 &
NONE &
\dots &
\phn\phn41.86784  &
$-$93.5733 &
1 \\

 & 
2002 Sep \phn9 & 
POL-S &
SCAMRR & 
\phn9 &
SPIRAL &
4 &
\phn\phn12.992\phn\phn  &
$-$108.88\phn\phn &
 \\


 & 
2003 Jun \phn8 & 
POL-S &
SCAMRR & 
\phn9 &
SPIRAL &
4 &
\phn\phn12.992\phn\phn  &
1.32\phn\phn &
\enddata
\tablenotetext{a}{POL-S: short wavelength polarizers 
(POL0S, POL120S, and POL240S); 
POL-L: long wavelength polarizers (POL0L, POL120L, and POL240L)}
\tablenotetext{b}{Total exposure time per polarizer.}
\tablenotetext{c}{The ORIENTAT header parameter refers to PA of the image +y axis (degrees E of N).}
\tablenotetext{d}{The ``Z'' prefix in the IRAS designation is given 
to indicate the fact that this object was found in the Faint Source 
Reject File in the {\iras} Faint Source Survey \citep{hrivnak99}.}
\tablenotetext{e}{PSF standard for our target sources.}
\tablenotetext{f}{PSF (and photometric) standard for the archived data.}
\tablerefs{%
1.\ \citet{hines00},
2.\ \citet{su03}}
\end{deluxetable*}

In addition to the four target sources, we also obtained 
data from a PSF reference star, HD 12088, since we concluded 
from our previous experience with WFPC2 that the best way to 
reduce the PSF effects from {\hst} data was to use observed 
PSF data \citep{ueta00}.
This high proper-motion star was selected as a PSF reference 
because 
(1) most unpolarized stars are known to be high proper-motion 
stars and 
(2) this star lies in the continuous viewing zone (CVZ) 
and can be observed in the same orbit as one of the 
target sources that is also in the CVZ without spending 
an additional orbit just for PSF observations.
The same instrumental set-up was used for the PSF reference 
observations.
We summarize the observing parameters in Table \ref{log}.

We are interested in the morphologies of the circumstellar 
shells revealed by the dust-scattered, polarized light.
Although polarimetric data have been acquired from evolved 
stars with NICMOS in the past, no study has been done to 
extract the spatial information from the data by alleviating 
the PSF effects of the bright central star. 
Therefore, we have analyzed all the archived NICMOS 
polarimetric data obtained from evolved stars (mostly 
done with NIC2) to study the shell structure.
Here, we included data from a PPN, {\iras} 16594$-$4656,
whose {\ipol} map provides more spatial information of 
the shell structure than previously reported \citep{su03}. 
{\iras} 16594$-$4656 was originally observed in a 
GO program 7840 (PI: S.\ Kwok), in which Camera 2 (NIC2; 
$19\farcs2 \times 19\farcs2$ field of view at $0\farcs075$ 
pixel$^{-1}$ scale) and long wavelength polarizers 
(POL0L, POL120L, and POL240L, covering 1.9 to 2.1 \micron) 
were used with the three-point spiral dither pattern.
Table \ref{log} also shows observing parameters for 
the additional sources. 

\subsection{Data Reduction}

We used the standard set of NICMOS calibration programs provided 
in the latest version (Version 3.1) of IRAF/STSDAS\footnote{%
STSDAS is a product of the Space Telescope Science Institute, 
which is operated by AURA for NASA} at the time of data reduction. 
The CALNICA calibration routines in STSDAS perform zero-read 
signal correction, bias subtraction, dark subtraction, detector 
non-linearity correction, flat-field correction, and flux calibration. 
After pipeline calibration, we realized that our data were 
affected by the ``pedestal effect'', a variable quadrant bias, 
and the ``Mr.\ Staypuft'', the amplifier ringing and streaking 
due to bright targets \citep{nicmoshandbook}. 
The pedestal effect was removed by first manually inserting the 
STSDAS task {\sl biaseq} in the middle of the CALNICA processes 
(before flat-fielding) and then employing the STSDAS task 
{\sl pedsub} after the CALNICA processes.
The former removes the non-linear components of the bias drift, 
while the latter takes care of the linear components.
To remove the ``Mr.\ Staypuft'' anomaly, we applied ``undopuft'' 
IDL routines provided by the STScI NICMOS group to the raw data 
before using any of the CALNICA routines. 
Any remaining stripes along the fast readout direction associated 
with ``Mr.\ Staypuft'' were subtracted by extracting the anomalous 
stripes from the block-averaged blank sky.

The calibrated, anomaly-cleaned frames of dithered images were 
then combined into a single image by applying the variable-pixel 
linear reconstruction algorithm (the STSDAS package {\sl dither}, 
Version 2.0; \citealt{dithering}).
This method would interlace (``drizzle'') each pixel in multi-point 
dithered frames according to the statistical significance of each 
pixel.
The drizzled images were sub-pixelized by a factor of 2 during the 
process: the reduced images would have pixel scales of $0\farcs0215$ 
pixel$^{-1}$ (NIC1) and $0\farcs0375$ pixel$^{-1}$ (NIC2).
Cosmic-rays were also removed by the drizzling algorithm.

The NICMOS arrays are slightly tilted with respect to the focal 
plane, and thus the NICMOS pixel scales along the two spatial 
axes of each camera are not identical.
Although its effect is small, pixels have to be properly rectified 
to accurately determine the orientation of the polarization vectors. 
Since the geometric distortion parameters have been found to be 
identical before and after SM3B, we simply applied the geometrical 
transformation within the drizzle process, using the geometric 
distortion coefficients provided by the STScI NICMOS group.

\newpage
\subsection{Derivation of the Stokes Parameters}

At the end of the processes described above, images
would have units of the default count rate for {\hst} 
data (DN sec$^{-1}$; ``DN'' = data number).
These count rates can be translated into appropriate 
physical units by means of photometric conversion 
factors, PHOTFLAM and PHOTFNU, before the Stokes 
parameters are derived from the data.
However, these factors depend on the actual detector 
sensitivities, which were altered after the NCS 
installation due to the 15 K raise of the NICMOS 
detector operating temperature.
Since post-SM3B photometric conversion factors 
for the polarizers were not available at the time 
of our analysis, we estimated them following 
the method described below.

First we retrieved archived NICMOS calibration data 
of an unpolarized standard star, BD$+$32$^{\circ}$3739, 
from CAL programs executed before and after SM3B 
(PIDs: 7692, 7958, and 9644).
Then, we reduced the data following the same procedure 
as our scientific data and performed aperture photometry 
on the calibration data using the {\sl apphot} task in 
the NOAO/DIGIPHOT package in IRAF.
From the measured count rates of the pre-SM3B data and 
the pre-SM3B photometric conversion factors (PHOTFNUs),
we computed the flux of the standard star through each
polarizer.
Finally, we obtained the post-SM3B PHOTFNUs from the 
derived flux of the standard star and the measured 
count rates of the post-SM3B data.
The measured count rates of the scientific data were 
then converted into Janskys.
Table \ref{photfnu} summarizes PHOTFNU values we used.

 \begin{deluxetable}{lccccc}[hb]
\tablecolumns{6} 
\tablewidth{0pt} 
\tablecaption{\label{photfnu}%
Photometric and Polarimetric Calibration Parameters} 
\tablehead{%
\colhead{} & 
\multicolumn{2}{c}{PHOTFNU (Jy sec DN$^{-1}$)} &
\colhead{} & 
\multicolumn{2}{c}{$t_{k}$} \\
\cline{2-3}
\cline{5-6}
\colhead{Polarizer} &
\colhead{Pre-SM3B} &
\colhead{Post-SM3B} &
\colhead{} & 
\colhead{Pre-SM3B} &
\colhead{Post-SM3B}} 

\startdata 

POL0S &
6.996~E$-$06 &
4.31~E$-$06  &
&
0.7766     &
0.7760     \\

POL120S    &
6.912~E$-$06 &
4.19~E$-$06  &
&
0.5946     & 
0.5935     \\

POL240S    &
6.914~E$-$06 &
4.15~E$-$06  &
&
0.7169     &
0.7181     \\

POL0L      &
7.626~E$-$06 &
6.17~E$-$06&
&
0.7313     &
0.8774     \\

POL120L    &
7.530~E$-$06 &
6.10~E$-$06&
&
0.6288     &
0.8381     \\

POL240L    &
7.517~E$-$06 &
6.04~E$-$06&
&
0.8738     &
0.9667    
\enddata
\end{deluxetable}

Using the photometric calibrated data for each polarizer,
we derived the Stokes parameters following the matrix 
inversion method \citet{hines00,nicmoshandbook}.
For the post-SM3B data, we used revised matrix coefficients 
provided by the STScI NICMOS group, of which only the 
$t_{k}$ coefficients (which are related to the throughput 
of the polarizers) have been updated from the pre-SM3B 
values.
The $t_{k}$ values are listed in Table \ref{photfnu}.

\begin{figure}[t]
\includegraphics[width=88mm]{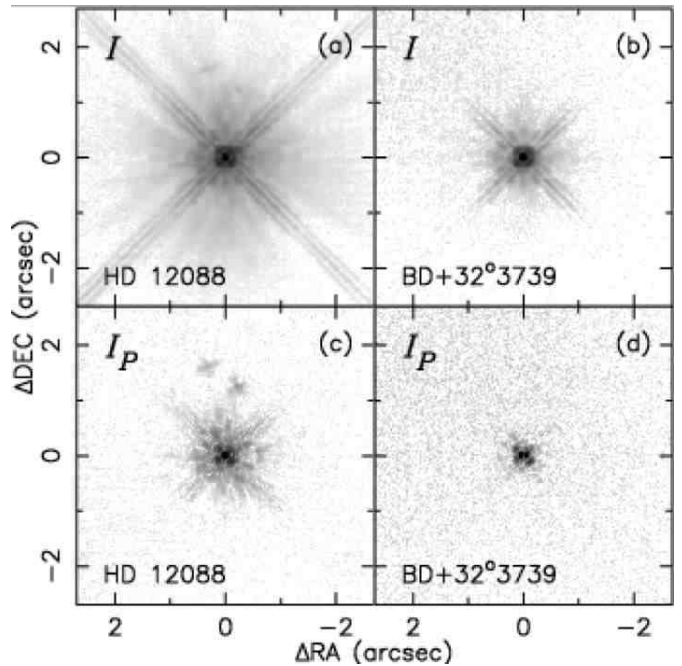}
\caption{\label{psf}%
The $I$ (top) and {\ipol} (bottom) maps of 
our PSF standard, HD12088 (left), and an 
unpolarized standard, BD$+$32$^{\circ}$3739
(right).
The maps are centered at the position of the star and the 
tickmarks indicate the RA and DEC offsets in arcsec.
The data +y direction points up and the data +x direction 
to the left: the PSF structure and polarizer ghost show 
at the same location.}
\end{figure}

\subsection{PSF Subtraction}

To measure correct polarization strengths, the PSF 
effects have to be removed.
Thus, we need to confirm the polarization free 
nature of our PSF standard star, HD 12088.
Fig.\ \ref{psf} shows the $I$ and {\ipol} maps of 
HD 12088 and of an archived polarimetric standard 
star, BD$+$32$^{\circ}$3739.
HD 12088 does not appear to be extended (besides 
the more pronounced PSF spikes and polarizer ghosts)
with respect to BD$+$32$^{\circ}$3739 in $I$ 
(Figs.\ \ref{psf}a and \ref{psf}b).
The {\ipol} map (Fig.\ \ref{psf}c) does not seem to 
indicate the presence of intrinsic polarization 
from HD 12088 other than the known polarizer ghosts 
appearing at ($0\farcs23, 1\farcs19$) and 
($-0\farcs16, 0\farcs93$) in comparison with
BD$+$32$^{\circ}$3739 (Fig.\ \ref{psf}d).
The excess {\ipol} structure is likely due to photon 
shot noise caused by the bright star and is restricted 
mainly to $\ale 0\farcs6$ from the star.
The measured polarization is
$1.5\%$ in a $0\farcs6$ diameter aperture (about 
$\times 3$ larger than the FWHM of the PSF)
and $2.8\%$ in pixels registering 
$> 10 \sigma_{\rm sky}$ in the $I$ map for HD 12088,
while that for BD$+$32$^{\circ}$3739 
is $1.8\%$ and $1.4\%$, respectively.
\citet{hines00} reported $\ale 1\%$ instrumental 
polarization.
Thus, our $\age 1\%$ polarization of BD$+$32$^{\circ}$3739
is likely due to systematics in the data reduction
procedure.
Slightly higher polarization of HD 12088 in the
$> 10 \sigma_{\rm sky}$ pixels is likely due to 
the polarizer ghosts and pixels affected by the PSF
spikes.
Therefore, we consider that polarization in HD 12088 is
negligible for our purposes in detecting polarization
much higher than a few \% and
that such small polarization would not affect our
interpretation of the data.

Closer inspection of Figs.\ \ref{psf}a and \ref{psf}b 
indicates that there is a companion object very close 
to HD 12088 at ($-0\farcs19, 0\farcs09$).
Although such a companion can be removed by combining 
images rotated by a multiple of $90\deg$ around the 
central star, it turned out that this operation would 
average out the asymmetric PSF structure to a degree 
that the PSF subtraction would not be effective.
Thus, we performed PSF-subtraction simply using the 
raw HD 12088 maps: the resulting PSF-subtracted maps 
would suffer from oversubtraction by this companion 
object.

For the PSF removal, we tried deconvolution using 
the Lucy-Richardson method (the {\sl lucy} task in 
IRAF) and the MCS method \citep{mcs}.
However, deconvolution did not yield satisfactory 
results because (1) the PSF effects were too severe 
and extensive to be removed and (2) the companion 
objects interfered with the deconvolution algorithm.
Thus, we resorted to a much simpler scale-and-shift 
subtraction approach.
The intensity scaling factor and shifts between the 
images were found by iteratively searching for the 
best values using the PSF spikes,
since we did not know {\it a priori} how much 
intrinsic nebular flux there is in addition to the 
stellar flux and the pixels over the central 
star were often susceptible to large photon shot 
noise.
It should be noted that the HD 12088 data are only 
effective for our target data obtained with NIC1 
polarizers after SM3B.
For PSF subtraction in pre-SM3B NIC2 data, we used 
the BD$+$32$^{\circ}$3739 data.

\begin{figure}[t]
\includegraphics[width=88mm]{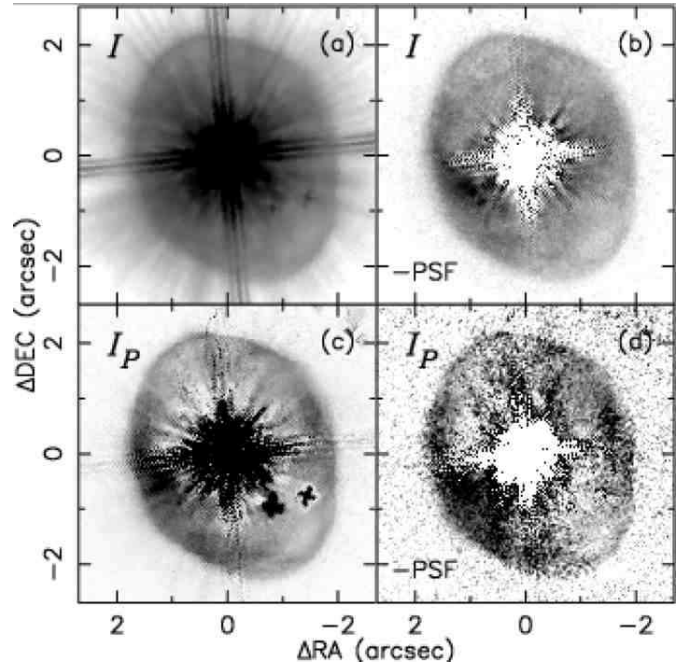}
\caption{\label{psfsub}%
The $I$ (top) and {\ipol} (bottom) maps of {\iras} 07134
derived from the raw (left) and PSF-subtracted (right) data.
The maps are in the standard orientation (N is up, E to the left)
and centered at the central star with tickmarks showing the 
RA and DEC offsets in arcsec.
Detailed shell structure can be seen in the PSF-subtracted $I$ 
map, and even better in the {\ipol} maps.
While the PSF-subtracted {\ipol} map shows almost artifact-free
structure of the shell, it is degraded from the low S/N of the
PSF standard data.}
\end{figure}

Fig.\ \ref{psfsub} demonstrates improvements 
gained from PSF subtraction.
The raw $I$ map (Fig.\ \ref{psfsub}a) is severely 
affected by the PSF, and it is nearly impossible 
to gain any spatial information about the nebulosity 
except that it is extended.
However, the PSF-subtracted $I$ map (Fig.\ \ref{psfsub}b) 
successfully reveals the {\sl internal} shell structure.
The raw {\ipol} map (Fig.\ \ref{psfsub}c) shows even 
more shell structure than the PSF-subtracted $I$ map:
polarized light seems to be concentrated at the 
periphery of the nebulosity with local brightness 
enhancements.
However, we now recognize the intrusive PSF spikes 
and polarizer ghosts at ($-0\farcs8$, $-0\farcs9$) 
and ($-1\farcs4$, $-0\farcs8$) due to photon shot 
noise caused by the extremely bright central star 
(Fig.\ \ref{psfsub}d) shows the PSF-subtracted {\ipol} 
map in which the shell structure is seen almost
artifact-free.
However, the quality of the image is gravely 
compromised by the S/N of the PSF standard data.
This is why data from the polarimetric standard star 
have to be carefully tailored in imaging polarimetry 
for optically thin circumstellar shells surrounding
a bright central source.
Unfortunately, we did not have an independent 
orbit for PSF observations.
Hence, our PSF data was unable to gain enough S/N 
to achieve optimum results from PSF subtraction.
In the following, we use raw {\ipol} data in order 
to take advantage of high S/N unless the polarizer 
ghosts pose serious problems in the data analysis.
As for the $P$ maps, we use raw $P$ data for 
displaying purposes, but measurements were done 
with the PSF-corrected data.

 \begin{deluxetable*}{lrrcccccccc}
\tablecolumns{11} 
\tablewidth{0pt} 
\tablecaption{\label{results}%
Summary of {\hst}/NICMOS Polarimetric Results} 
\tablehead{%
\colhead{} & 
\multicolumn{2}{c}{Measured Coord.\ (J2000)\tablenotemark{a}} &
\colhead{} &
\multicolumn{3}{c}{Flux} &
\colhead{$\left<P\right>$\tablenotemark{c}} &
\multicolumn{3}{c}{{\ipol} Structure} \\
\cline{2-3}
\cline{5-7}
\cline{9-11}
\colhead{Source} &
\colhead{RA} &
\colhead{DEC} &
\colhead{} &
\colhead{Band\tablenotemark{b}} &
\colhead{$I$ (mJy)} &
\colhead{{\ipol} (mJy)} &
\colhead{($\%$)} &
\colhead{Morphology} &
\colhead{Size ($\arcsec$)} &
\colhead{P.A.\ ($^{\circ}$)\tablenotemark{d}}} 

\startdata 

IRAS 07134 & 
07 16 10.27 & 
$+$09 59 48.5 & 
&
POL-S &
9300\tablenotemark{e} & 
2000\tablenotemark{e} &
$55\pm16$ &
Elliptical/Round Hollow Shell &
$4.8 \times 4.0$ &
\phn25 \\

IRAS 06530 & 
06 55 31.80 & 
$-$02 17 28.3 & 
& 
POL-S &
\phn380 & 
\phn\phn67 &
$47\pm15$ &
Elliptical Shell $+$ Bipolar Cusp &
$2.7 \times 1.0$ &
\phn20 \\

IRAS 04296 & 
04 32 56.95 & 
$+$34 36 13.1 & 
& 
POL-S &
\phn340 & 
\phn\phn60 &
$40\pm12$ &
Quadrupolar \\
&
&
&
&
&
&
&
&
Elliptical (extension 1) &
$2.1 \times 0.7$ & 
\phn26 \\
&
&
&
&
&
&
&
&
Elliptical (extension 2)&
$3.5 \times 0.5$ &
\phn99 \\

IRAS 02229 & 
02 26 41.79 & 
$+$62 21 22.2 & 
& 
POL-S &
4800 & 
1200 &
$52\pm19$ &
Elliptical &
$2.1 \times 1.3$ & 
\phn59 \\

IRAS 16594 & 
17 03 10.04 & 
$-$47 00 27.0 &
& 
POL-L &
\phn620 & 
\phn\phn66 &
$39\pm16$ &
Elliptical $+$ Protorusions &
$5.0 \times 2.2$  &
104
\enddata
\tablenotetext{a}{Of the $I$ peak location.}
\tablenotetext{b}{%
POL-S: $0.8 - 1.3\micron$, centered at $1.1\micron$;
POL-L: $1.89 - 2.1\micron$, centered at 2.05\micron}
\tablenotetext{c}{Mean $P$ and its standard deviation.}
\tablenotetext{d}{Degrees E of N.}
\tablenotetext{e}{The measured $I$ flux value is three times higher 
than the previously observed value; this is in part due to photon
persistence.  See text for details (\S 4).}
\end{deluxetable*}

\section{Results}
 
\begin{figure*}[p]
\centerline{%
\includegraphics[width=115mm]{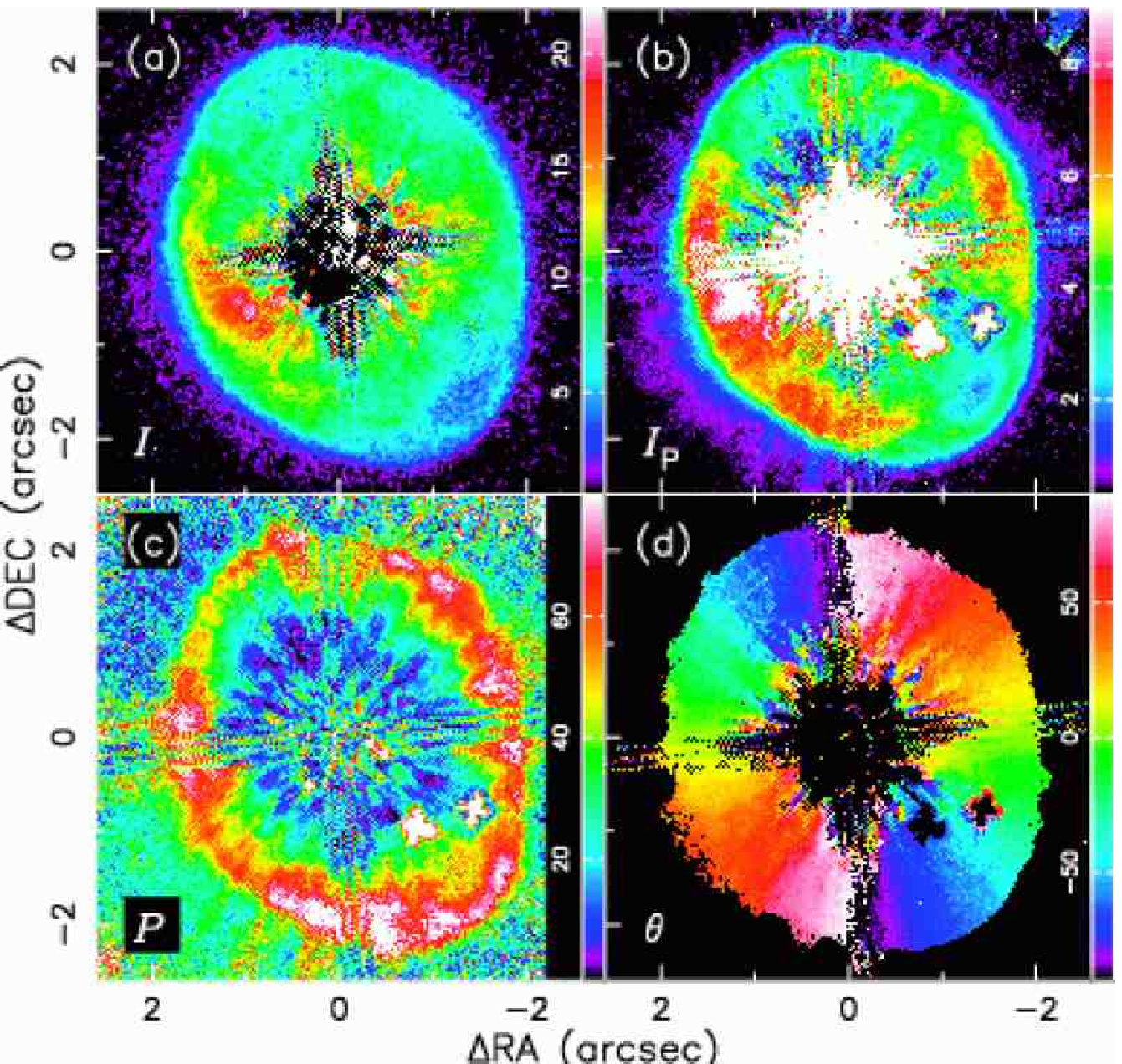}}
\caption{\label{07134}%
Polarimetric maps of {\iras} 07134$+$1005:
(a) the total intensity ($I$\/),
(b) polarized intensity (\ipol), 
(c) polarization strengths ($P$\/),
and (d) polarization PA ($\theta$\/), 
respectively from left to right, top to bottom.
The maps are in the standard orientation (N is up, 
E to the left) and centered at the $\theta$ center 
with tickmarks showing the RA and DEC offsets in 
arcsec.
The wedges indicate the scale of the image tone:
mJy arcsec$^{-2}$ in $I$ and {\ipol}, percentage 
in $P$, and degrees east of north in $\theta$ 
(i.e. PA $0^{\circ}$ means the polarization vector, 
which is perpendicular to the scattering plane, 
is oriented in the N-S direction). 
In displaying the $\theta$ map we used pixels which 
register S/N of $> 10$ $\sigma_{\rm sky}$ in the $I$ map.}
~\\
\centerline{%
\includegraphics[width=115mm]{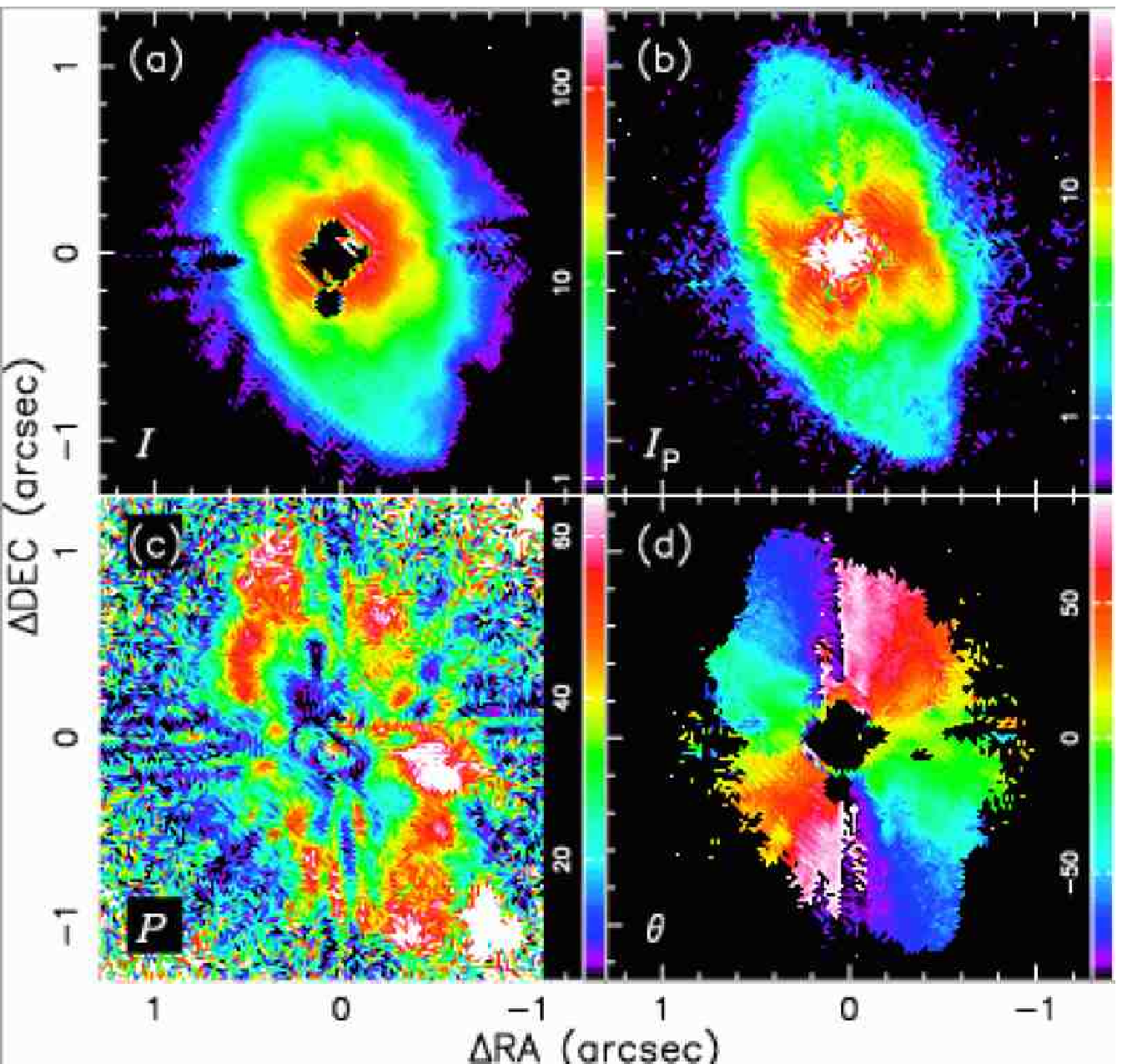}}
\caption{\label{06530}%
Polarimetric maps of {\iras} 06530$-$0213.
The display convention is the same as Fig.\ \ref{07134}.}
\end{figure*}

\begin{figure*}[p]
\centerline{%
\includegraphics[width=115mm]{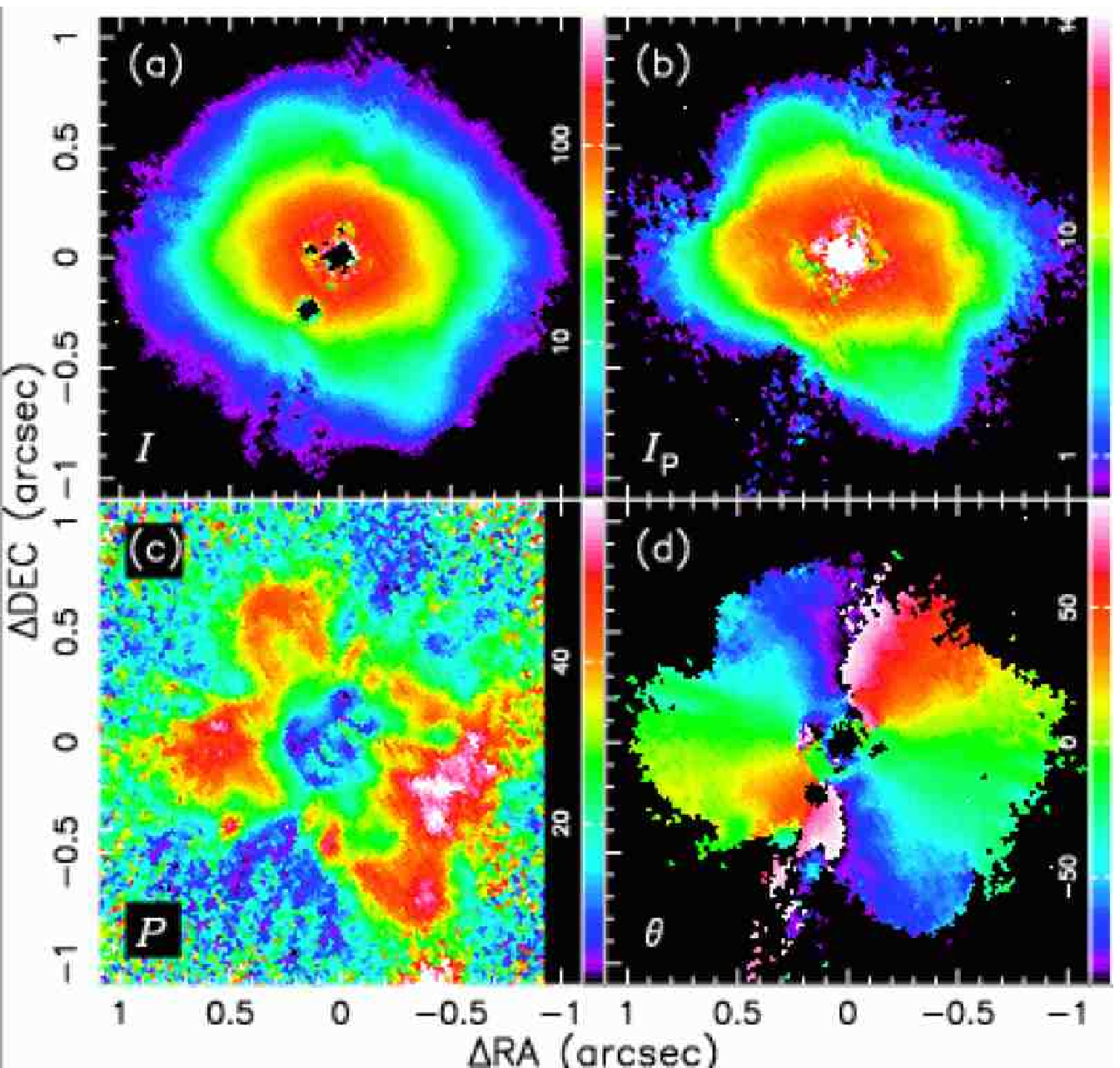}}
\caption{\label{04296}%
Polarimetric maps of {\iras} 04296$+$3429.
The display convention is the same as Fig.\ \ref{07134}.}
~\\
\centerline{%
\includegraphics[width=115mm]{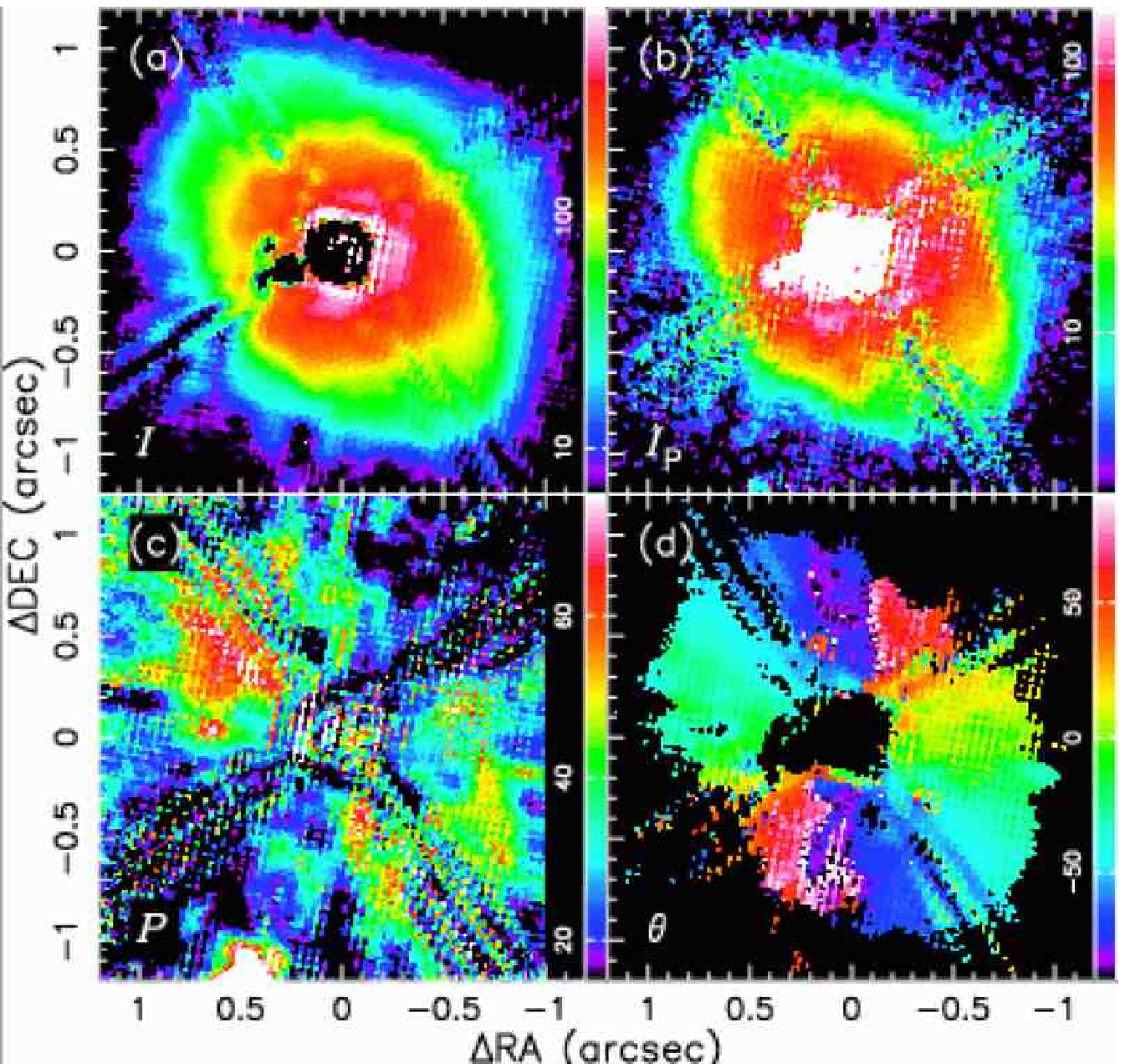}}
\caption{\label{02229}%
Polarimetric maps of {\iras} (z)02229$+$6208.
The display convention is the same as Fig.\ \ref{07134}.}
\end{figure*}

\begin{figure*}[t]
\centerline{%
\includegraphics[width=115mm]{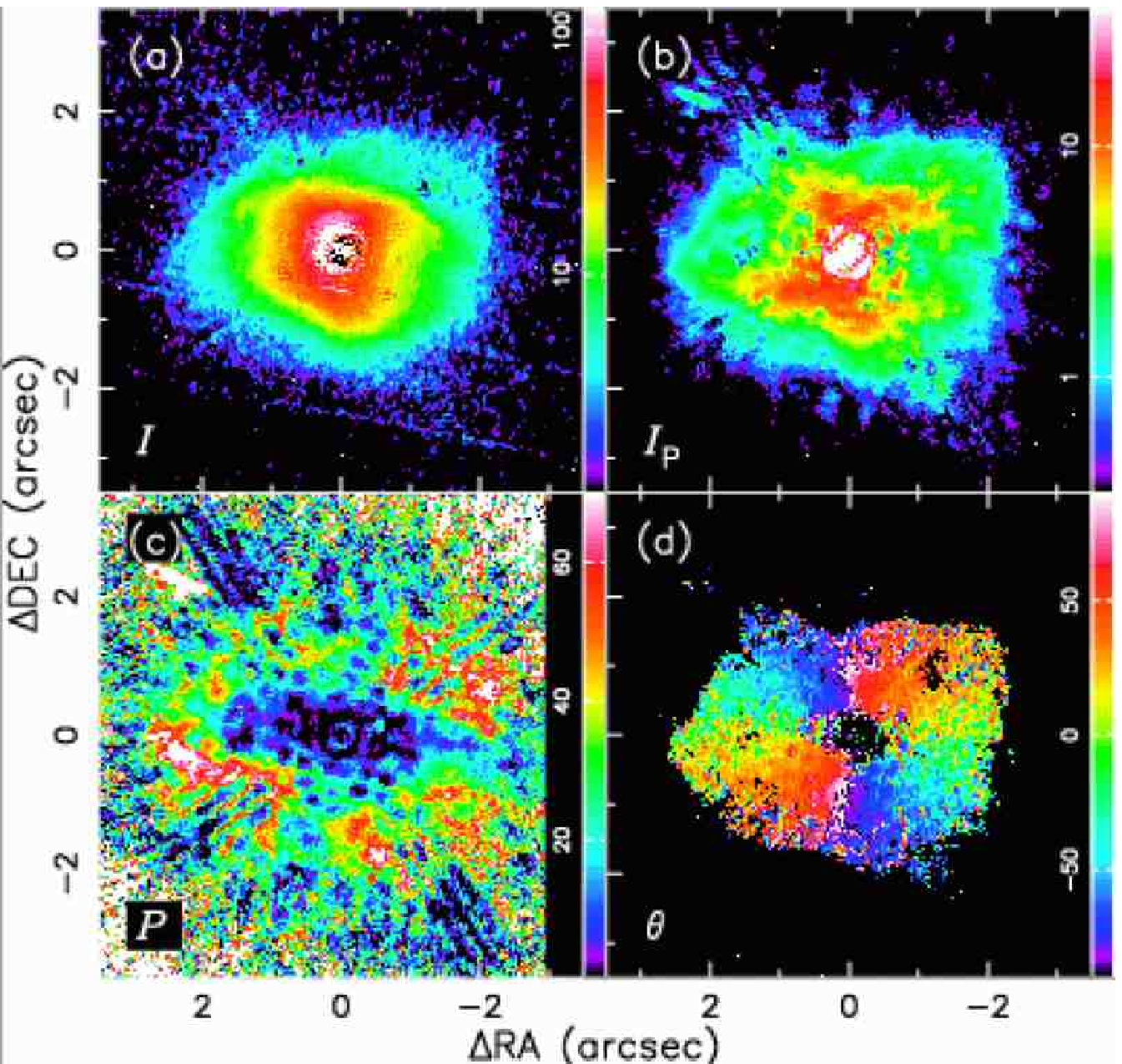}}
\caption{\label{16594}%
Polarimetric maps of {\iras} 16594$-$4656.
The display convention is the same as Fig.\ \ref{07134}.}
\end{figure*}

In Figs.\ \ref{07134} to \ref{16594}, we present 
the polarimetric maps of five PPNs in a uniform 
format.
The PSF-subtracted $I$ and {\ipol} (top left and 
right, respectively) maps reveals the distribution 
of scattering medium in these circumstellar shells.
The $P$ map (bottom left) represents the distribution 
of {\ipol} with respect to $I$, augmenting the 
$I$ and {\ipol} maps from a different point of view.   
The $\theta$ map shows the orientation of the 
polarization vectors by the ``rainbow wheel'' 
pattern.
We opt to display the $\theta$ maps this way
since the high resolution quality of the data can be 
compromised by inevitable rebinning in making 
conventional vector diagrams.
In addition, the pixels affected by image anomaly
(by the ghosts and shot noise near the central star)
have been masked out in the $\theta$ maps.

In Table \ref{results} we summarize the results of 
the observations including the measured coordinates 
of the object, $I$ and {\ipol} fluxes, mean $P$ and 
its standard deviation, and descriptions of {\ipol} 
structure.
The coordinates listed are the observed location of 
the $I$ peak, which coincides with the {\ipol} peak 
and the $\theta$ center if the shell is sufficiently 
optically thin.
The fluxes are determined by integrating the surface 
brightness over the shell where pixels register more 
than one $\sigma_{\rm sky}$.
Although sky emission has been subtracted in calibration,
we removed any residual ``sky'' emission if the sky 
value determined in an annulus around the object 
registers more than three $\sigma_{\rm sky}$.
The mean $P$ and its standard deviation have been 
determined by using pixels that registered more than 
10 $\sigma_{\rm sky}$.
The {\ipol} structure is described by the overall
shape, dimensions (typically major and minor axis 
lengths), and PA measured east of north.

The accuracy of the photometric results depends 
on the quality of the PHOTFNU values used in our 
analysis (Table \ref{photfnu}).
The measured $I$ fluxes are all consistent with 
the known photometric values, given the difference 
in the filter profiles between $J$/$K$ and short/long 
wavelength polarizers, except for {\iras} 07134.
Our {\iras} 07134 observations have yielded the $I$ 
flux of 9.3 Jy, which is more than a factor of three 
higher than recent measurements of 2.9 Jy 
(m$_{\rm J} = 6.8$; e.g.\ \citealt{ueta03a}).
While this source suffers from the polarizer ghosts, 
they are known to cause only less than $1 \%$ in 
brightness of the primary source \citep{hines00}.
Thus, the ghosts alone could not have introduced 
this inconsistency.
Given that our PHOTFNU have yielded reasonably consistent
photometric results for the other three targets (within 
$50\%$ difference), our estimates of the PHOTFNU values
do not seem to have caused systematic errors.
The most likely source of this inconsistency seems to be
the photon persistence.
The data for {\iras} 07134 are affected by severe photon 
persistence, and the affected pixels ($\sim 0\farcs4$ of
the star) can remain affected even two dither positions 
later (i.e., four exposures or more later).
The persistent signal decay can interfere with the 
non-linearity correction algorithm in the pipeline
calibration, leading to inaccurate count rates.
We have not, however, attempted to improve the accuracy
of our photometric measurements, since
(1) absolute photometric calibration is not possible 
without properly calibrated PHOTFNU values,
(2) only relative calibration among the three polarizers 
is important in deriving the Stokes parameters ($Q$ and $U$\/),
and
(3) only the vicinity of the star ($\sim 0\farcs4$) 
is affected by the photon persistence.

With the ($P$, $\theta$\/) data set, we see highly 
centrosymmetric nature of polarization in all PPNs. 
We can use the polarization 
vectors to backtrack the position of the illumination 
source, i.e., the central star.
The center of the vector pattern was derived by minimizing 
the sum of the square of the distance between the vector
position and the vector pattern center.
In this analysis, we used vectors that are in the pre-defined 
annulus centered at the presumed pattern center.
We assumed the $I$ peak to be the pattern center, 
and iterated the process by using the updated center position
until the shift between the previous and current centers
becomes smaller than the numerical accuracy of the analysis.
The vector pattern center was found to coincide with the 
$I$ peak position in all cases:
the results are summarized in Table \ref{centerposition}.
Thus, these nebulae - SOLE-toroidal PPNs - are indeed 
optically thin and illuminated by the central star
located at where the $I$ peak is.

 \begin{deluxetable}{lc}
\tablecolumns{2} 
\tablewidth{0pt} 
\tablecaption{\label{centerposition}%
Offsets between the illumination source and the $I$ peak} 
\tablehead{%
\colhead{} & 
\colhead{Offsets} \\
\colhead{Source} &
\colhead{(arcsec)}} 

\startdata 

IRAS 07134 &
$0.08 \pm 0.27$ \\

IRAS 06530 &
$0.04 \pm 0.20$ \\

IRAS 04296 &
$0.06 \pm 0.17$ \\

IRAS 02229 &
$0.12 \pm 0.31$ \\

IRAS 16594 &
$0.06 \pm 0.14$ 

\enddata
\end{deluxetable}

\section{Individual Sources}

In this section, we describe the individual source 
structure, including the 3-D aspects of it as revealed 
by these polarimetric images.

\subsection{IRAS 07134$+$1005 (HD 56126)}

Polarization observations of {\iras} 07134 (HD 56126) 
are presented in Fig.\ \ref{07134}.
The PSF-subtracted $I$ map (Fig.\ \ref{07134}a) shows a 
slightly elliptical nebula ($4\farcs8 \times 4\farcs0$ 
at PA $25^{\circ}$), which clearly possesses some 
internal structure previously unrecognized in the near-IR.
The bulk of surface brightness is concentrated in the 
region close to the minor axis of the nebula 
(PA $115^{\circ}$), with the eastern side being brighter 
and spatially more extended than the western side.
There is an apparent brightness peak on the east side
of the nebula at ($1\arcsec$, $-0\farcs5$), whereas the 
west side does not show any local brightness peak.
At the northern and southern end of the nebula, 
there is significantly less surface brightness.
There appears to be a filamentary structure 
along the periphery of the nebula, delineating the 
elliptical tips in the low surface brightness region 
(also seen well in the grayscale image, Fig.\ \ref{psfsub}b).

\begin{figure}[t]
\includegraphics[width=88mm]{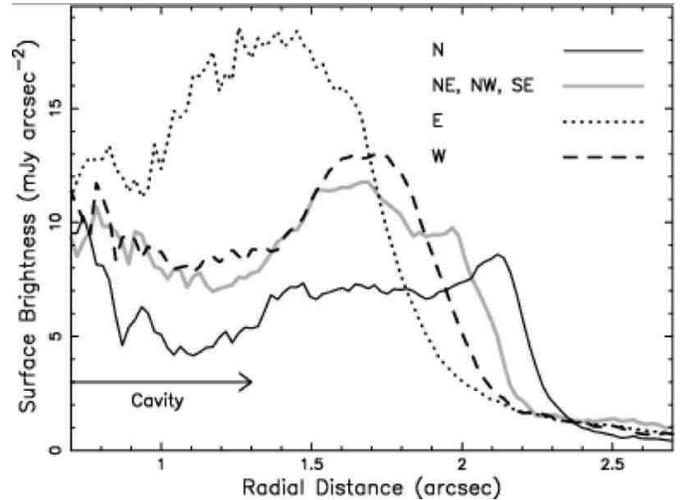}
\caption{\label{07134cut}%
Polarized brightness profiles of {\iras} 07134:
N cut (at PA $25^{\circ}$; black solid line), 
NW-NW-SE cut (average of cuts at intermediate PAs 
$70^{\circ}$, $160^{\circ}$, and $340^{\circ}$; gray solid line),
E cut (at PA $115^{\circ}$; dotted line), and
W cut (at PA $295^{\circ}$; dashed line).}
\end{figure}

In the {\ipol} map (Fig.\ \ref{07134}b), we can identify 
at least two regions of enhanced surface brightness on 
the east and west side of the nebula with the local peaks 
located at ($1\arcsec$, $-0\farcs5$) and 
($-1\farcs5$, $0\farcs5$).
The east peak shows stronger and more extended brightness 
distribution than the west peak.
The surface brightness of the east peak is $18$ mJy 
arcsec$^{-2}$ which is about 1.5 times brighter than the 
western counterpart (see the profiles in Fig.\ \ref{07134cut}).
The region of enhanced surface brightness 
($\age 5$ mJy arcsec$^{-2}$) extends from
PA $50^{\circ}$ to $205^{\circ}$ on the eastern side, 
and from $250^{\circ}$ to $350^{\circ}$ (with
an apparent gap at $320^{\circ}$) on the western side.
These brightness-enhanced regions are connected by 
the lower brightness ($\sim 3$ mJy arcsec$^{-2}$) region 
at the N-S elliptical tips of the nebula, in which 
filamentary structures outline the edge of the tips 
(more apparently in {\ipol} than in $I$).
Incidentally, the interior region encircled by this ``rim''
region shows weak surface brightness 
($\ale 3$ mJy arcsec$^{-2}$, in the region 
$\ale 1\farcs2 - 1\farcs3$ from the center).
Overall, the bulk of {\ipol} appears radially confined 
to the nebula periphery beyond about $1\farcs5$ from the
central star in all directions.
This is particularly seen well in the northern half of 
the nebula where surface brightness of the interior region
($\ale 1\farcs5$ from the central star) becomes very 
small (almost null) without any contamination by the 
polarizer ghosts 

In the $P$ map (Fig.\ \ref{07134}c), we immediately 
see that the high $P$ ($\age 20\%$) regions occur near 
the periphery of the nebula beyond about $1\farcs5$ 
from the central star in all directions: the mean 
polarization strength is $55\%$.
The high $P$ regions correspond to the high {\ipol} 
regions.
However, in the $P$ map, there is not so much of a
difference in the polarization strength at the elliptical 
tips and at the eastern/western edges of the nebula as 
in the {\ipol} map.
The northern tip shows somewhat weaker polarization strengths 
than the southern tip: this is consistent with slightly 
higher $I$ in the northern tip and almost equal {\ipol}
at the both tips.
The $\theta$ map (Fig.\ \ref{07134}d) illustrates the 
polarization PA by the image tone.
This particular $\theta$ map shows an almost perfect 
``rainbow wheel'' pattern by the uniform and symmetric 
gradation of the image tone in the azimuthal direction,
which depicts the highly centrosymmetric nature of the 
polarization.

The past observations of this PPN found a 
slightly elliptical nebula via dust-scattered 
star light in the optical \citep{ueta00} and 
its two-peaked core structure via thermal 
dust emission in the mid-IR 
\citep{meixner97,dayal98,jura00,kwok02}.
The observed morphology has been thought to 
represent an almost edge-on ellipsoidal 
(slightly prolate) shell with an equatorial 
density enhancement (i.e., torus) that results 
in limb-brightened two-peak core emission in 
the mid-IR.
This interpretation has been corroborated by 
a 2-D radiative transfer model of dust 
emission \citep{meixner97,meixner04}.
The present observations have revealed the 
toroidal structure of the shell for the first 
time in dust-scattered star light in the 
near-IR at more than a factor of two better 
resolution than the past mid-IR imaging.
Moreover, the polarization characteristics 
of the shell confirm that it is optically 
thin at 1 $\micron$ having the central star 
as the illumination source.
This further strengthens the edge-on torus 
interpretation, in which such density 
structure would manifest itself as two 
limb-brightened peaks only when the 
shell is sufficiently {\sl optically thin}.

\begin{figure}[t]
\includegraphics[width=88mm]{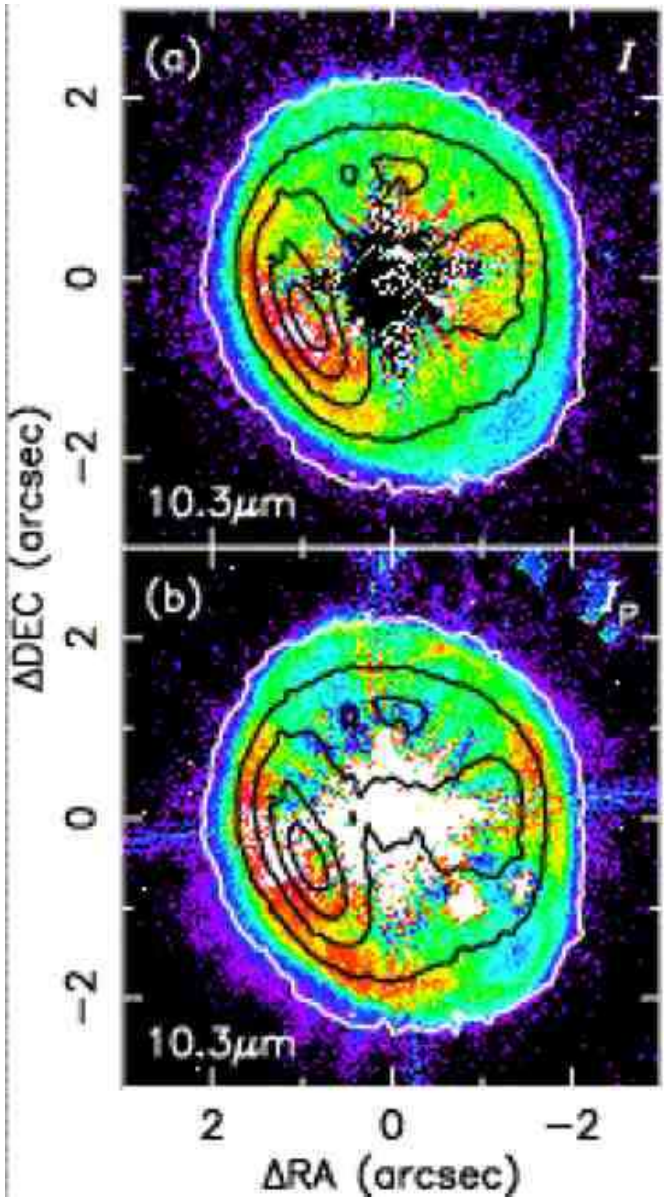}
\caption{\label{07134mir}%
Spatial relationship between dust-scattered light 
($I$ [top] and {\ipol} [bottom] maps in color) and 
thermal dust emission
(10.3 $\micron$ contours at 10, 30, 50, 70, and $90\%$;
\citealt{kwok02}).
The display convention is the same as Fig.\ \ref{07134}.}
\end{figure}

In Fig.\ \ref{07134mir}, we compare the shell 
structure seen in dust-scattered near-IR light 
(color) and in thermal dust emission at 10.3 
$\micron$ (contours; from \citealt{kwok02}).
Both data show similar brightness distribution where 
the eastern side is stronger and covers a larger 
spatial extent:
even the surface brightness gap on the western side 
at PA $320^{\circ}$ is seen in both of the {\ipol}
and 10.3 $\micron$ maps.
Thus, we confirm that the imbalance of brightness 
distribution is real, and so is the isolated emission 
blob north of the central star seen in the mid-IR.
Geometrically, this blob appears to be part of the 
western edge of the toroidal structure that has been 
broken off for some reason.

There has been a question of whether this imbalance 
of mid-IR peak strengths is due to density or temperature 
effects of dust grains.
If the dust temperature were the cause for the peak 
strength imbalance, we would not have seen the same
imbalance in the {\ipol} map.
Polarization maps are sensitive to scattered light,
whereas thermal dust emission maps are sensitive to
warm dust.
Thus, the {\ipol} distribution does not necessarily 
have to follow the distribution of thermal mid-IR 
emission.
Together with the mid-IR data,
our images suggest that there is simply 
more dust grains in the eastern side of the shell 
than the western side.
Recent CO observations have shown a similar
morphology \citep{meixner04}, corroborating that 
there is more matter (dust and gas alike) on the 
eastern side of the shell in this object.

Meanwhile, there are apparent differences between
the {\ipol} and mid-IR morphologies which can be 
easily understood by the differences in the nature 
of light arising from dust grains.
One difference is that the {\ipol} and mid-IR peaks 
do not spatially coincide: the {\ipol} peaks are 
found more towards the edge of the shell than the 
mid-IR peaks.
In general, mid-IR emission arises from the warmest 
($\sim 100$ to $200$ K) dust grains located near 
the inner edge of the shell.
Thus, we would expect the mid-IR peak at the edge 
of the inner cavity where the line of sight 
traverses the longest distance in the warmest dust.
However, due to the curvature of the surface of 
the inner cavity, the peak mid-IR position tends 
to be found somewhat closer to the central star.
On the other hand, {\ipol} becomes the strongest 
at the inner boundary and decreases in the radially 
outward direction assuming a radially decreasing 
density profile (e.g. $\propto r^{-2}$), because 
scattering geometry dictates the behavior of 
scattered light.
Another difference is that the {\ipol} peaks are 
more extended along the nebula edge than the 
mid-IR peaks.
This is simply because scattering can occur as long as
there is enough incident light and scattering medium.
Thus, the {\ipol} map has unveiled the shell structure 
in the outer shell where the region of enhanced 
density is extended well into the high latitude 
part of the ellipsoidal shell.

As a PPN, {\iras} 07134 is expected to have an 
inner cavity generated by the cessation of mass 
loss at the end of the AGB phase.
The emission structure of the mid-IR and CO maps 
(e.g. \citealt{kwok02,meixner04}) is consistent 
with the presence of such a cavity.
If the shell of {\iras} 07134 is hollow and
has a radially decreasing density structure, 
then we expect that
(1) the highest {\ipol} occurs at the inner 
edge of the shell and 
(2) $P$ should radially increase due to the 
geometrical effect of scattering angles (confined 
closer to $90^{\circ}$) and becomes the highest
at the outer edge of the shell.
Our {\ipol} and $P$ maps do show these characteristics
exactly as expected.
Although it is still possible that the contamination
by the unpolarized component of the PSF artificially
lower $P$ in the central region, 
the high $P$ ($\age 20\%$) region is restricted near
the outer edge of the shell beyond the PSF.
This high $P$ regions occupy the same portions of the
shell as the high {\ipol} regions, forming a ``ring''
structure at the rim of the shell.
Therefore, the shell of {\iras} 07134 most likely 
possesses an inner cavity, and the part of the shell 
probed by scattered light represents a hollow spheroid.

To better illustrate the spatial variation of surface 
brightness, we have made various cuts in the {\ipol}
map.
In Fig.\ \ref{07134cut}, we show profiles along the
northern major axis (N cut; PA $25^{\circ}$;
solid black line),
eastern minor axis (E cut; PA $115^{\circ}$;
dotted line),
western minor axis (W cut; PA $295^{\circ}$;
dashed line),
and
intermediate PAs (NE-NW-SE cut; PAs $70^{\circ}$,
$160^{\circ}$, and $340^{\circ}$;
solid gray line).
These profiles are derived from a linear cut of 
10-pixel width (or the median of multiple cuts, in 
the case of the NE-NW-SE cut).
We do not include the S and SW cuts because of the
contamination by the polarizer ghosts.

These cuts show a similar profile: there is a steep 
outer edge representing dust pile-up, which surrounds
the high brightness region of the main shell, and
the brightness steeply falls down to a relatively flat, 
plateau region.
All cuts show the inner plateau inward of 
$\ale 1\farcs3 - 1\farcs4$, except for the E cut 
with plateau of $\ale 0\farcs9$.
We interpret that this inner ``plateau'' profile is 
due to the inner cavity created by a precipitous 
drop in the mass loss rate at the end of the AGB phase.
In the context of polarization, the presence of the
inner cavity means the absence of the scattering 
medium near the plane of the sky, resulting in an
abrupt decrease of {\ipol} and $P$, thereby forming 
an inner boundary of the shell.
It is, therefore, the direct observational evidence 
for the presence of such an inner cavity in PPNs. 

\subsection{IRAS 06530$-$0213}

\begin{figure}[t]
\includegraphics[width=88mm]{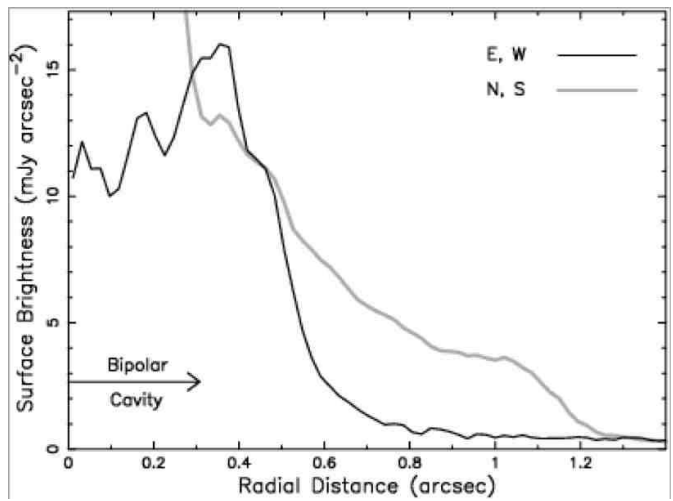}
\caption{\label{06530cut}%
Polarized brightness cuts of {\iras} 06530:
N-S cut (average of N and S profiles at PA $20^{\circ}$ and 
$110^{\circ}$; gray solid line) and 
E-W cut (average of cuts along lines perpendicular to 
the major axis $0\farcs43$ N and S of the equatorial plane; 
black solid line).}
\end{figure}

The polarization maps of {\iras} 06530 are 
displayed in Fig.\ \ref{06530}.
The PSF-subtracted $I$ map (Fig.\ \ref{06530}a) 
shows a highly elliptical shape of the nebula 
($2\farcs4 \times 1\farcs1$ at PA $20^{\circ}$), 
in which the low brightness ($\sim 5$ mJy arcsec$^{-2}$) 
elliptical tips extend beyond the barrel-shaped 
region of higher surface brightness ($\age 15$ 
mJy arcsec$^{-2}$).
The {\ipol} map (Fig.\ \ref{06530}b) reveals a 
cusp-like (or a sideway {\it x}) structure within 
the barrel region.
Such structure typically indicates the swept-up
walls of the bipolar cavities.
Thus, {\iras} 06530 is likely a near edge-on, 
highly prolate ellipsoidal shell showing a rather 
bipolar nature in the low latitudinal region.
Fig.\ \ref{06530cut} shows the polarized brightness 
cuts of the shell made along the major axis 
(N-S cut; N-S averaged, thick gray line) and the 
lines parallel to the minor axis at 
$0\farcs43$ N and S of the central star 
(E-W cut; E-W averaged, solid black line).
The N-S cut demonstrates the elongated {\ipol} 
structure with a gentle slope at the edge, while
the E-W cut shows a peak at around $0\farcs34$ 
that defines a bipolar cavity wall.
The polarization characteristics seen in the $P$ 
and $\theta$ maps (Fig.\ \ref{06530}c and 
\ref{06530}d, respectively) are very similar to 
those of {\iras} 07134 (Fig.\ \ref{07134}).
$P$ is stronger near the edge of the shell while
it is absent in the central region even beyond 
the PSF.
The polarization pattern is very much centrosymmetric
as seen from the almost perfect rainbow wheel
pattern.

Thus, {\iras} 06530 is a highly prolate spheroidal 
shell, which very likely has an inner cavity 
as in {\iras} 07134.  
Although the shell is optically thin, there is 
sufficiently high concentration of dust grains 
at the equatorial region so that the bipolar 
cusp structure is seen in {\ipol}.
The total optical depth is not large enough to 
induce the full extinction of star light expected in 
typical bipolar PPNs (e.g., {\iras} 17150$-$3224; 
\citealt{kwok98,su03}).
This may be why we do not observe a low {\ipol}
``hole'' in the central region of the shell as
we saw in {\iras} 07134.
However, we do see a possible inner cavity
in the $P$ map.

\subsection{IRAS 04296$+$3429\label{discuss04296}}

In Fig.\ \ref{04296} we present data from {\iras} 04296.
The PSF-subtracted $I$ map (Fig.\ \ref{04296}a) shows a 
quadrupolar nebula 
(roughly $1\arcsec \times 1\farcs5$) with an east-west 
elongated core having round protrusions towards PA 
$26^{\circ}$ and $206^{\circ}$.
The {\ipol} map (Fig.\ \ref{04296}b) successfully unveils 
the slanted {\sl X} shape of the nebula more clearly than 
the $I$ map.
The nebula's {\sl X} shape is due to the presence of two 
axes of elongation.
One of the elongations is oriented at PA $26^{\circ}$ 
and shows a relatively well-defined elliptical shape 
($2\farcs1 \times 0\farcs7$; extension 1). 
The other elongation, oriented at PA $99^{\circ}$, is 
fainter with its surface-brightness-limited ends smeared 
out in the background ($3\farcs5 \times 0\farcs5$; 
extension 2).
These elongations are not oriented perpendicular to each 
other.
Although \citet{sahai99} reported that extension 2 is
not straight with a $5^{\circ}$ shift between the 
eastern and western tips from the WFPC2 data, we are unable
to confirm this in our data due to the confusion
by the PSF spike nearly aligned with this extension.
While the {\ipol} structure is consistent with the one seen 
in the previous WFPC2 data \citep{ueta00} the central star 
appears more prominently in the near-IR data, since 
the central star (the nebula) is brighter (fainter) in the
near-IR.
This is why the $I$ map does not clearly show the {\sl X} 
structure as in the WFPC2 images.

\begin{figure}
\includegraphics[width=88mm]{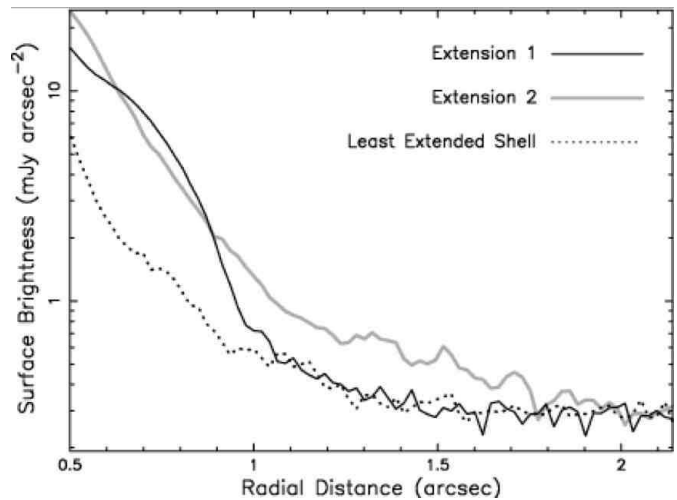}
\caption{\label{04296cut}%
Surface brightness radial profiles of {\iras} 04296:
of the extension 1 (at PA $26^{\circ}$; black solid line), 
of the extension 2 (at PA $99^{\circ}$; gray solid line), 
and of the least extended part of the shell (the average
of the cuts at PA $63^{\circ}$ and 
$153^{\circ}$; dotted line).}
\end{figure}

Fig.\ \ref{04296cut} displays {\ipol} radial profiles of 
the extensions to better present the extent of the 
structures.
These profiles are constructed by taking linear cuts of 
10-pixel width along the extensions and averaging the 
values at both ends.
For comparison, we also show the ``least extended'' shell 
profile, created from linear cuts at PA $63^{\circ}$ and 
$153^{\circ}$ (in-between directions of the extensions).
The extension 1 profile (black solid line) shows {\ipol} 
excess in the region close to the central star, but it 
suddenly falls to the flux level similar to the least
extended profile (black dotted line) at around $1\arcsec$.
On the other hand, the extension 2 profile exhibits {\ipol}
excess as far out as $\sim 1\farcs8$ before it gradually 
falls down to the background level.

The $P$ map (Fig.\ \ref{04296}c) uncovers the structure 
of the two extensions even better.
In extension 1, strongly polarized light ($\age 30\%$) 
is concentrated near the periphery of the elongation, 
especially at the tips, whereas weakly polarized light 
fills the cavity surrounded by the high polarization 
region.
Because of the compactness of the nebula, we can not 
rule out the possibility that the central region registers 
low $P$ values due to the residual unpolarized component 
from the central star.
The $P$ structure of extension 1 resembles that of
the previous two sources.
However, in extension 2, the weak $P$ cavity is not 
very well-defined partly due to confusion by the presence 
of falsely high polarization caused by the PSF spikes.
Nevertheless, high polarization is observed as far out 
as $1\arcsec$ from the central star. 
The $\theta$ map (Fig.\ \ref{04296}d) exhibits a general 
centrosymmetric pattern of an optically thin shell, with 
only a marginally shallower gradient over the extensions.

Based on the optical morphology of the nebula, 
\citet{sahai99} interpreted extension 1 to be a 
bounded disk that collimated outflows manifesting 
themselves as extension 2.
Our polarization data, however, do not support
this disk interpretation.
The {\ipol} shape of extension 1 does not 
support a geometrically thin disk structure.
If extension 1 is a thin disk with its plane inclined 
$24^{\circ}$ with respect to the line of sight, the 
spatial distribution of the disk material in the plane
of the sky will be extremely restricted.
This would result in strong scattering in a 
geometrically narrow region in the sky, and 
the {\ipol} map will likely be very narrow in 
the direction perpendicular to the extension.
However, this is not the case.
The fact that extension 1 appears elliptical in 
{\ipol} strongly suggests that the distribution 
of scattering matter is elliptical in the plane 
of the sky, implying a prolate spheroidal 
structure of the shell.
The $P$ morphology of extension 1 shows a possible 
low $P$ cavity surrounded by the region of strong 
polarization at the nebula edge.
Thus, extension 1 is very likely a prolate spheroid,
possibly with an inner cavity.
In terms of polarization characteristics, there is no 
significant difference in both extensions except for 
the sharpness of the tip structure.
In extension 1, we see the highest $P$ at the tips
of the elongation.
However, in extension 2, the highest $P$ does not
arise from the tips.
As the profiles indicate in Fig.\ \ref{04296cut}, 
extension 2 is extended out to about $1\farcs8$.
This may suggest that extension 2 is also a hollow
prolate spheroid, but is inclined with respect to 
the plane of the sky so that the region of high $P$
occurs where the spheroidal shell intersects with 
the plane of sky and not at the tips.

\subsection{IRAS (Z)02229$+$6208\label{discuss02229}}

We show the polarization maps of {\iras} 02229 in Fig.\ 
\ref{02229}.
The PSF-subtracted $I$ map (Fig.\ \ref{02229}a) shows an 
elliptically extended nebula ($2\farcs1 \times 1\farcs3$ 
at PA $45^{\circ}$), which is consistent with the previous 
WFPC2 images \citep{ueta00}.
The surface brightness distribution is such that the
southwestern  
tip is slightly brighter than the northeastern tip.
The {\ipol} map (Fig.\ \ref{02229}b) exhibits the same 
elliptical shape of the nebula.
Unlike the $I$ map, the {\ipol} map does not show any 
apparent spatial difference in the polarized surface 
brightness distribution.
Although the low brightness edge ($\sim 50$ mJy arcsec$^{-2}$) 
is elongated towards PA $45^{\circ}$ (nearly aligned with 
a PSF spike), the high surface brightness core ($> 200$ 
mJy arcsec$^{-2}$) appears to be elongated towards a 
slightly different direction (PA 60$^{\circ}$).

The $P$ map (Fig.\ \ref{02229}c) appears quite differently
with respect to other $P$ maps: strong polarization is not 
concentrated near the periphery in this nebula.
Instead, we see a band of low polarization ($\ale 30\%$) 
in the middle of the shell aligned with PA 150$^{\circ}$, 
which separates the region of medium polarization 
($\sim 30 - 40\%$) on the southwest side and the region 
of high polarization ($\age 40\%$) on the northeast side.
The $\theta$ map (Fig.\ \ref{02229}d) shows a generally 
centrosymmetric pattern.
However, the gradient of the vector angle seems to be steep 
in the low polarization band, and shallow in the elliptical 
tips of the shell.
This indicates that more vectors in the elliptical tips 
are aligned parallel to the low $P$ band which most likely
represents the equatorial plane.
Note also that the orientation of the low polarization 
band is not perpendicular to the direction of the elongation
of the shell, but to that of the core elongation 
(PA $60^{\circ}$).
The polarization characteristics of this object is very 
distinct with respect to those of the previous objects.

The peculiar $P$ map can be understood in terms of the inner 
structure and inclination of the shell.
The presence of the low $P$ band suggests that dust density 
is more equatorially enhanced within a rather geometrically 
narrow region in this object.
If the near side of this inner torus is tilted towards the 
northeastern direction, the southwestern side of the shell 
(i.e., the near side of the ellipsoidal shell) appears more 
illuminated by the direct star light in our viewing angle.
This is consistent with the $I$ map showing the brighter 
southwestern tip than the northeastern tip (Fig.\ \ref{02229}a).
However, assuming a spheroidal density distribution of the shell
there will be no difference in terms of the amount 
of scattering medium in the plane of the sky, and hence,
the {\ipol} brightness will be the same on both sides 
of the shell (Fig.\ \ref{02229}b). 
Since $P$ is a ratio of {\ipol} to $I$, we would see lower
(higher) $P$ on the southwestern (northeastern) side of the 
shell.  
Therefore, {\iras} 02229 seems to be similar to {\iras} 
06530 with a spheroidal shell with a relatively stronger 
equatorial density enhancement.  
However, the shell orientation of {\iras} 02229 is most 
likely more pole-on compared to {\iras} 06530.

\begin{figure}[h]
\includegraphics[width=83mm]{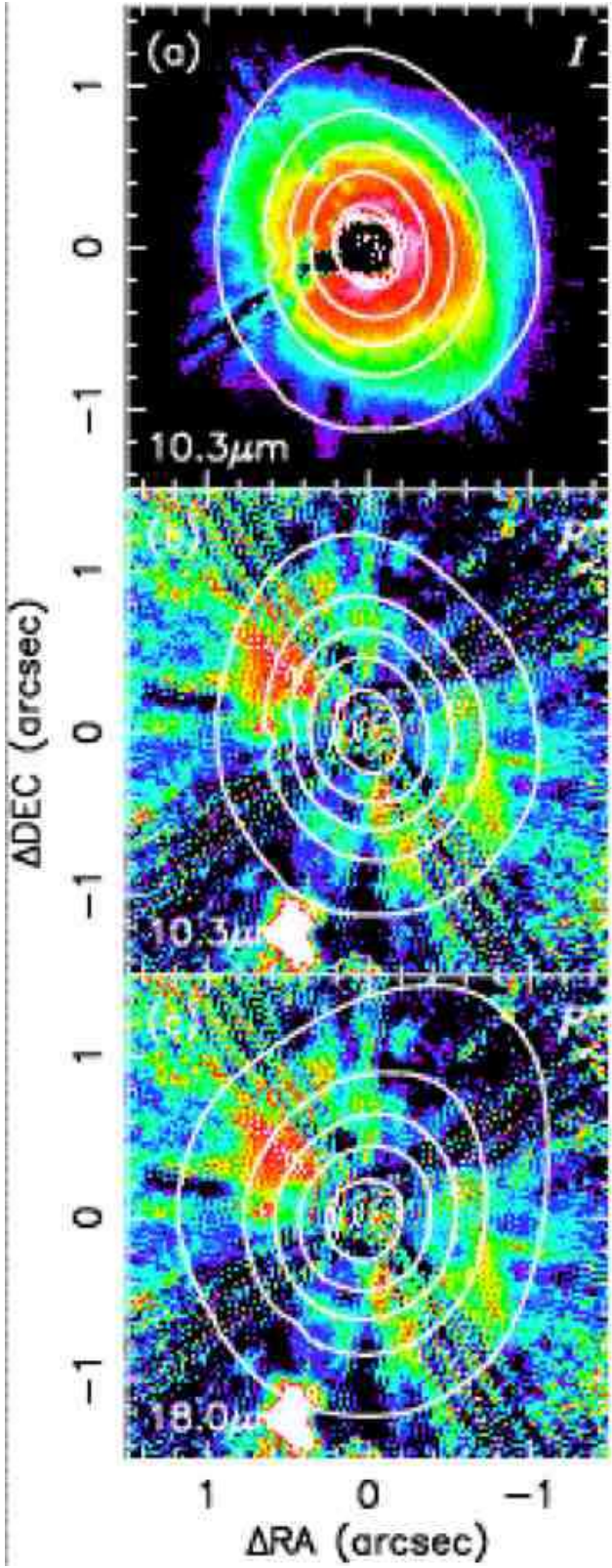}
\caption{\label{02229mir}%
{\iras} 02229: Spatial relationship between 
dust-scattered light and thermal dust emission
(color {\ipol} map and 10.3 $\micron$ contours; [a]),
the polarization strength and thermal dust emission 
(color $P$ map and 10.3 and 18.0 $\micron$ contours;
[b] and [c], respectively).
The mid-IR data are from \citet{kwok02}, and
the contours are at 10, 30, 50, 70, and $90\%$.
The display convention is the same as Fig.\ \ref{07134}.}
\end{figure}

{\iras} 02229 was marginally resolved in the mid-IR 
in our previous imaging survey \citep{meixner99}, and 
has recently been observed at the Gemini North \citep{kwok02}.
Fig.\ \ref{02229mir} shows the {\ipol} and $P$ maps overlaid
with 10.3 and 18.0 $\micron$ contours.
10.3 $\micron$ emission is extended with PA of $20^{\circ}$, 
which is off by $25^{\circ}$ with respect to the $I$ 
elongation (Fig.\ \ref{02229mir}a).
We have also seen a slight shift of PA between the outer
and inner $I$ structure ($45^{\circ}$ and $60^{\circ}$,
respectively).
Assuming these parts of the shell with different PAs
represent distinct portions of the shell structure, 
this shift of PA may indicate that the inner torus is 
precessing/rotating in the counterclockwise direction.
Such precession/rotation of the inner torus has already
been suspected in a SOLE-toroidal PPN, HD 161796 
\citep{gledhill03}.

The low $P$ band also does not seem to show strong spatial
correlation to the 10.3 $\micron$ emission region 
(Fig.\ \ref{02229mir}b).
However, the 18.0 $\micron$ map seems to be elongated in
the direction of the low $P$ band (Fig.\ \ref{02229mir}c).
\citet{kwok02} suspected that the 18.0 $\micron$ map 
may have suffered from variable sky conditions,
If {\iras} 02229 has geometrically narrow density distribution 
in the innermost region of the shell,
it is possible that the mid-IR continuum emission would
be elongated along the equatorial plane as we see in
Fig.\ \ref{02229mir}c.
If this geometry causes the 18.0 $\micron$ elongation, 
then this raises a question as to why the 10.3 $\micron$
emission, another map of continuum, is not elongated
as the 18.0 $\micron$ map.
The 10.3 $\micron$ map, in fact, is morphologically very
similar to the 11.7 and 12.5 $\micron$ maps that represent
unidentified IR feature emission.
Although \citet{kwok02} concluded that 10.3 $\micron$
emission represented continuum, comparison of the {\sl ISO}
spectrum with the mid-IR filter profiles (see Fig.\ 4 in
\citealt{kwok02}) may indicate that the 10.3 $\micron$
filter does capture some emission due to the same
unidentified IR feature emission that caused elongation 
in the 11.7 and 12.5 $\micron$ images.
It is also interesting to note that the 20.8 $\micron$
map (representing the unidentified ``21 $\micron$'' feature)
shows somewhat different elongation with respect to
other maps.
Thus, it may be that the 18.0 $\micron$ map is the true
continuum emission map that reflects the disk-like dust 
distribution of the shell, and that other maps are 
different due to distinct dust species responsible for 
feature emission at other wavelengths.
Further investigation at higher resolution,
preferably spatial spectroscopy, is necessary 
to determine if distinct spatial distribution of different 
dust species can result in such changes of morphology.

\subsection{IRAS 16594$-$4656}

We present the archived 2 $\micron$ polarization data 
of {\iras} 16594$-$4656 that have been reanalyzed to 
mitigate the PSF effects (Fig.\ \ref{16594}).
The PSF-subtracted $I$ map (Fig.\ \ref{16594}a) reveals 
the inner structure of the nebula that is clearly 
elongated in the east-west direction ($5\farcs0 \times 2\farcs2$ 
at PA $81^{\circ} - 261^{\circ}$; $\age 7$ mJy arcsec$^{-2}$).
The {\ipol} map (Fig.\ \ref{16594}b) uncovers the structure
of the shell, in which we see the bipolar cusp structure,
similar to the one in {\iras} 06530 (Fig.\ \ref{06530}b),
corresponding to the main $I$ elongation.
We do not see any elliptically elongated tips surrounding 
the cusp structure as in the case for {\iras} 
06530.
The tips of the cusp are more elongated in this object 
and appear to delineate the wall of the bipolar cavities.

The $P$ structure resembles those with a hollow shell
(Fig.\ \ref{07134}c, \ref{06530}c, and probably \ref{04296}c), 
in 
which there is a region of low polarization in the middle 
of the shell encircled by a region of high polarization.
In {\iras} 16594, however, the region of high polarization
does not seem to completely surround the central low $P$
region: it is mainly found southeastern and northwestern 
ends of the shell.
Moreover, it appears that the $P$ map also possesses a
low $P$ band structure in PA $167^{\circ}$, which 
is similar to the one in {\iras} 02229 
(Fig.\ \ref{02229}c).
The $\theta$ map is very much centrosymmetric.

{\iras} 16594 has been observed with {\hst} many times 
in the past.
WFPC2 images have shown the object's multi-polar reflection 
nebula that has three extensions on each side
(e.g.,  \citealt{hks99}).
These extensions form oppositely pointing pairs in the
directions of PA $40^{\circ} - 220^{\circ}$,
$60^{\circ} - 240^{\circ}$, and $80^{\circ} - 260^{\circ}$.
The length of the elongation decreases in the 
counterclockwise direction, while the surface brightness increases.
Thus, the optical structure has been interpreted as
oppositely directed material ejected episodically
from a rotating source.
Our PSF-subtracted $I$ and {\ipol} images show that 
the inner shell structure is aligned with the elongation 
of PA $80^{\circ} - 260^{\circ}$, as has been seen only 
faintly in heavily PSF affected NICMOS broad to medium 
band images \citep{su03}.
Hence, the episodic ejection interpretation is consistent
with the nebula's inner shell structure.

The toroidal nature of the innermost shell has recently 
been exposed by mid-IR imaging, in which 
the presence of the limb-brightened peaks in the emission 
core has proven that the shell has an equatorial density 
enhancement along PA $170^{\circ}$ \citep{gh04}.
The elongation of the inner shell is therefore aligned 
with the symmetric axis of the torus.
Thus, it seems likely that material ejection is presently 
channeled into the directions of the inner shell elongation
by the torus or by some collimation mechanism(s) that can 
generate such outflows and equatorial density enhancement.

The orientation of the torus is spatially coincident with 
the low $P$ band seen in our data (Fig.\ \ref{16594}c).
This confirms our interpretation of the presence of low 
$P$ band as a manifestation of dust density enhancement 
along the equatorial plane of the system (see Fig.\ \ref{02229}c
and discussion associated with it).
\citet{su03} interpreted that the low $P$ around the central 
star was due to an inclined dust torus at an ``intermediate''
angle.
However, the $P$ map of {\iras} 16594 does not suggest
an inclination angle as large as {\iras} 02229 in which
the imbalance of $I$ due to inclination resulted in the 
imbalance of $P$ on the opposing sides of the shell.
Thus, the orientation of the torus in {\iras} 16594 is 
more likely close to edge-on.
A dust emission model with our 2-D radiative transfer 
code has indicated the inclination angle of roughly 
$75^{\circ}$ with respect to the line of sight \citep{ueta04}.

\section{Discussion}

\subsection{3-D Shell Structure via ({\ipol}, $P$)}

Our main objective in the present study is to 
investigate the structure of {\sl optically thin}
PPNs by means of imaging polarimetry.
We have made use of the ({\ipol}, $P$\/) data set 
to achieve our goal instead of the more conventional 
($P$, $\theta$\/) data set with which 
{\sl optically thick} regions of the circumstellar 
shells are probed through the way polarization 
vectors are aligned with respect to the equatorial 
plane (e.g.\ \citealt{wh93,su03}).

In radially decreasing density distribution 
typical of PPNs, {\ipol} radially decreases.
So, {\ipol} becomes the strongest at the inner 
edge of the shell if the shell has an inner cavity 
as in the case of PPNs.
In the optically thin regime where single scattering 
dominates, $P$ in general becomes the strongest 
at the outer edge of the shell because of the
geometrical effect.
With the ({\ipol}, $P$\/) data set,
we have successfully detected both inner 
and outer edges of the shell in {\iras} 07134. 
The data have shown that the polarized
surface brightness distribution encircles 
the inner cavity, forming a complete ``ring''.
In terms of scattering geometry, such 
brightness distribution indicates a hollow
spheroid.
The data have also shown equatorial 
enhancement in the material distribution.

Our previous studies of the PPN structure 
independently confirmed (1) equatorially 
enhanced (toroidal) 
dust distribution in the innermost region of
the shell by mid-IR imaging \citep{meixner99,ueta01},
(2) the presence of an elliptical shell 
surrounding the central torus \citep{ueta00}
by optical imaging, and
(3) an equatorially enhanced hollow spheroid
(a combination of the above structures)
embedded in a spherically symmetric outer 
shell would explain all the morphological 
characteristics by numerical modeling
\citep{meixner02}.
With imaging polarimetry, we have been able 
to observationally prove that the PPN structure 
is a hollow spheroid with a built-in equatorial 
enhancement.

Standard imaging data show only the structure 
of the circumstellar shells projected to the 
plane of the sky: the structural information 
along the line of sight is degenerate.
However, the ({\ipol}, $P$\/) data set can retain 
the 3-D properties of these shells because the 
{\ipol} and $P$ strengths depend on the scattering 
geometry within the shells.
Thus, we can effectively probe the structure of 
PPNs, by extracting their 3-D information.
In addition, we have been able to determine 
the detailed geometry of our target sources
from their polarization properties:
{\iras} 04296 has a quadrupolar shell and does 
not harbor a disk with collimated outflows 
and {\iras} 02229 is oriented in a rather inclined 
direction with respect to us.

\subsection{Morphological Classification Scheme of PPNs}

\citet{gledhill01} introduced a classification scheme for 
their imaging polarimetry of PPNs based on the {\ipol}
morphology and other polarization properties.
These categories are {\sl Shells}, {\sl Bipolars}, 
and {\sl Core-dominated}.
Their study confirmed the SOLE-toroidal vs.\ 
DUPLEX-core/elliptical bifurcation found among PPNs 
and supported the idea that 
the bifurcation originated from the varying degrees of 
optical thickness of the shell.
Our target PPNs are all {\sl Shells}
based on their polarization properties (that is, they are
equivalent to SOLE-toroidals).
However, our high resolution ({\ipol}, $P$\/) data set 
has shown a range of morphologies among these 
{\sl optically thin} PPNs, from a detached shell 
structure ({\iras} 07134; Fig.\ \ref{07134}b, Fig.\ \ref{07134cut})
to a bipolar cusp structure ({\iras} 16594; Fig.\ \ref{16594}b), 
and even the mixture of the two ({\iras} 06530; Fig.\ \ref{06530}b,
Fig.\ \ref{06530cut}).
This suggests that even for {\sl optically thin}
PPNs there are multitudes of optical depths that 
their shells can assume.

As pointed out by \citet{gledhill01}, the division 
of the SOLE-DUPLEX bifurcation is not clearly 
defined.
These morphological classes are the both ends of
a spectrum in the domain of optical depth.
A given PPN can have any optical depth 
in this continuous distribution of optical depth,
and therefore, can assume any morphology along 
this SOLE-DUPLEX spectrum.
The $P$ maps of {\iras} 02229 and {\iras} 16594 
(Figs.\ \ref{02229}c \& \ref{16594}c) have also
hinted that inclination angles and highly geometrically
thin equatorial density enhancements can leave 
characteristic signature in the resulting morphology. 
Thus, we have demonstrated the robustness of 
our data, and our results further strengthen the 
suggestion that the optical depth of the shell plays 
a major role in determining the PPN morphology
with an added complexity from the actual geometry
of the equatorial enhancement and the inclination
of the object.
In order to quantitatively understand the PPN morphology
in the {\sl optically thin} regime, 
we need more scattering/polarization models of optically
thin shells with consideration of inclination angles.

\subsection{Structure of Superwind}

 \begin{deluxetable*}{lcllrcccrrc}
\tablecolumns{11} 
\tablewidth{0pt} 
\tablecaption{\label{dynamic}%
Dynamical Properties of the Sources} 
\tablehead{%
\colhead{} & 
\colhead{} & 
\colhead{$R_{\rm sw}$} & 
\colhead{$R_{\rm in}$} & 
\colhead{$D$} & 
\colhead{} & 
\colhead{$v_{\rm exp}$} & 
\colhead{} & 
\colhead{$t_{\rm sw}$} & 
\colhead{$t_{\rm dyn}$} & 
\colhead{} \\
\colhead{Source} &
\colhead{Direction} &
\colhead{(\arcsec)} &
\colhead{(\arcsec)} &
\colhead{(kpc)} &
\colhead{Ref.} &
\colhead{(\kms)} &
\colhead{Ref.} &
\colhead{(yrs)} &
\colhead{(yrs)} &
\colhead{Comments}} 

\startdata 

IRAS 07134 &
Pole (N) &
2.2 & 1.3 & 2.4 & 1 & 10.5 & 2 &
990 &
1430 &
\\
 &
Pole (S) &
2.5 & 1.3\tablenotemark{a} & 2.4 & 1 & 10.5 & 2 &
1300 &
1430 &
\\
&
Equator (E) &
1.8 & 0.9 & 2.4 & 1 & 10.5 & 2 &
990 &
990 &
Prolonged mass loss? \\
&
Equator (W) &
2.0 & 1.3 & 2.4 & 1 & 10.5 & 2 &
780 &
1430 &
\\

IRAS 06530 &
Pole &
1.35 & 0.9\tablenotemark{b} & 6.7\tablenotemark{c} & 3 & 14 & 3 &
1000 &
2100 &
Bipolar cavity \\
&
Equator &
0.5 & 0.35\tablenotemark{d} & 6.7\tablenotemark{c} & 3 & 14 & 3 &
350 &
800 &
Bipolar cusp \\

IRAS 04296 &
Extension 1 &
1.05 & 0.6 & 4.0 & 4 & 12 & 5 &
720 &
960 &
Co-existing flows? \\
&
Extension 2 &
1.75 & 0.6\tablenotemark{e} & 4.0 & 4 & 12 & 5 &
1800 &
960 &
Co-existing flows? \\

IRAS 02229 & 
Pole &
1.05 & \dots & 2.2\tablenotemark{f} & 6 & 13 & 6 &
850\tablenotemark{f} &
0\tablenotemark{f} &
\\
&
Equator &
0.65 & \dots & 2.2\tablenotemark{f} & 6 & 13 & 6 &
530\tablenotemark{f} &
0\tablenotemark{f} &
\\

IRAS 16594 &
Pole &
1.05 & 0.6\tablenotemark{b} & 2.2 & 7 & 16 & 8 &
220 &
1400 &
Precessing/Rotating Torus\\
&
Equator &
1.1 & 0.7\tablenotemark{d} & 2.2 & 7 & 16 & 8 &
260 &
460 &
Bipolar Cusp

\enddata
\tablecomments{%
$R_{\rm sw}$: superwind shell ({\ipol}) size;
$R_{\rm in}$: inner radius;
$D$: distance;
$v_{\rm exp}$: expansion velocity;
$t_{\rm sw}$: duration of superwind mass loss;
$t_{\rm dyn}$ dynamical expansion time}
\tablerefs{%
1. \citet{hony03},
2. \citet{meixner04},
3. \cite{hrivnak03},
4. \citet{meixner97},
5. \citet{omont93},
6. \citet{reddy99},
7. \citet{vds03},
8. \citet{loup90}}
\tablenotetext{a}{Assumed to be the same as in Pole (N).}
\tablenotetext{b}{Estimated from the $P$ profile.}
\tablenotetext{c}{2.8 times as far as {\iras} 07134,
provided the objects have the same luminosity
\citep{hrivnak03}.}
\tablenotetext{d}{Measured at $0\farcs7$ above and
below the equatorial plane.}
\tablenotetext{e}{Assumed to be the same as in Extension 1.}
\tablenotetext{f}{The lower limit.}
\tablenotetext{g}{Assuming no cavity.}
\end{deluxetable*}

PPNs form as a result of stellar mass loss along 
the AGB (c.f., \citealt{kwok93,balick02}).
The innermost shells are created by stellar material 
that has been ejected in the most recent mass loss, 
and thus, represent the latest AGB mass loss history.
In other words, the observed shells likely embody
the {\sl superwind} shells \citep{renzini81,iben83}.
According to the standard scenario for the AGB 
mass loss, superwinds are enhanced versions of
AGB winds that occur during the final stage of 
the AGB evolution (corresponding to the thermal
pulsing AGB phase; \citealt{iben95} for a 
thorough review).
Hydrodynamical simulations combined with detailed
stellar evolution models have shown that thermal 
pulses lead to enhanced mass loss events 
\citep{schroder99} and the formation of detached 
dust shells \citep{steffen98,steffen00}.
Therefore, we can use the observed shell
structure to glimpse the history of superwind 
mass loss at the end of the AGB phase.
Here, we will continue our discussion using 
{\iras} 07134 data that shows the density 
distribution in the greatest detail.

{\iras} 07134 has a rather well-defined outer
edge at 2\farcs0 to 2\farcs4 from the central 
star as seen in the $I$ and {\ipol} images 
(Fig.\ \ref{07134}a and \ref{07134}b) and 
surface brightness profiles (Fig.\ \ref{07134cut}).
At the edge, the signal registers at least 
10 $\sigma_{\rm sky}$, and hence, the edge is 
most likely density bounded.
The local surface brightness peak at the northern 
edge of the shell (Fig.\ \ref{07134cut}; i.e.,
the filamentary structure in the $I$ and {\ipol} 
maps, Figs.\ \ref{07134}a and \ref{07134}b) 
likely represents a pile-up of matter at the 
interface between the faster dust-driven wind 
and the slower shock-driven wind co-existing 
in the AGB shell seen in the model calculations
\citep{steffen00}.
Thus, the observed shell in dust-scattered star 
light very likely embodies the shell created by 
superwind, 
especially by the wind associated with the last 
thermal pulse.
Using the distance of 2.4 kpc \citep{hony03}
and the wind velocity of 10.5 {\kms} 
\citep{meixner04}, the edge represents matter
ejected 2000 to 2400 years ago.
We also note the presence of inner cavity of 
about 1\farcs4 radius, which is most likely 
due to the precipitous decrease of mass loss 
at the end of the AGB phase and defines the 
{\sl physically detached} nature of the PPN.
The width of the superwind shell is about 
0\farcs8.
So, the final mass loss has continued for about 
800 years after the edge matter was ejected,
and the shell has been expanding for about 
another 1400 years since the mass loss ceased
at the end of the AGB phase.
This duration of the final mass loss is 
consistent with theoretically derived 
duration of superwind mass loss 
\citep{bloecker95}.

The surface brightness profiles along the 
equatorial directions are evidence for 
distinct histories of mass loss experienced 
by the different equatorial regions of the 
shell.
The observed shell structures suggest that 
the equatorial density enhancement does not 
occur symmetrically.
There is larger amount of dust grains on the 
eastern side than the western side.
Assuming the constant mass loss velocity 
of 10.5 {\kms} on both sides of the shell, 
this imbalance could have resulted from 
prolonged superwind mass loss experienced by 
the eastern side:
while the western superwind terminated about 
1400 years ago, the eastern superwind continued 
another 400 years.
Alternatively, the eastern side may not have 
been moved as far away from the central star 
as the western side if the wind momentum is
equal on both sides.

The presence of the inner cavity confirms the 
physically detached nature of the shell in 
{\iras} 07134.
However, other objects do not show the inner 
cavity as clearly as {\iras} 07134, especially
in {\ipol}.
These shells instead show the bipolar cusp
structure around the central star.
Since the cusp structure suggests the presence 
of matter in the inner cavity, the presence of 
the cavity in these shells may not have been 
seen clearly in {\ipol} as a ``hole''.
Their $P$ structure is, nevertheless, very similar
to that of {\iras} 07134 
(Fig.\ \ref{06530}c for {\iras} 06530, 
Fig.\ \ref{04296}c for {\iras} 04296,
and, Fig.\ \ref{16594}c for {\iras} 16594).
Thus, we tentatively suggest their marginally 
hollow nature.
As for the cause of the cusp structure (and the
presence of matter in the inner cavity), it is
not very clear.
However, it may be intrinsic to PPNs that have
optically thicker shells.
That is, such cusp structure within the inner 
cavity may be {\sl required} to form as PPNs 
develop optically thicker shells.
The cusp structure may also be indicative
of the presence of a rather flattened torus, 
which can participate in forming highly elongated
shell as we see in these objects.
However, the degree to which this shell 
elongation mechanism works seems to be 
relatively low in these objects, and thus,
these nebulae do not possess the prototypical 
bipolar morphology.

The presence of a single inner shell in {\iras} 
16594 indicate that the multiple shell structure 
seen in the optical has been created one elongation
at a time while the central region precesses/rotates.
However, the quadrupolar structure of
{\iras} 04296 (Fig.\ \ref{04296}) suggests
two elliptical elongations that can co-exist.
Using the distance of 4 kpc \citep{meixner97} 
and the CO outflow velocity of 12 \kms, we 
estimate that the superwind which shaped 
extension 1 was initiated about 1580 years ago.
The absence of the pile-up edges in extension 
2 raises an important question about its origin.
If the wind velocity along extension 2 is 
similar to that along extension 1, the wind 
that shaped extension 2 must have been initiated 
another 1580 years earlier than the
beginning of extension 1 formation.
If the shaping of extension 2 was initiated 
about the same time as the shaping of extension 
1 by the onset of superwind, the wind velocity 
along extension 2 must have been about 24 \kms.
If the amount of dust in the two extensions 
is different (e.g. \citealt{sahai99}), two winds 
with the same velocity could generate elongations 
with differing length.
Either case, two elongations seem to have
been generated concurrently at least for some time.
Thus, there are likely multiple channels to create 
multi-polar shell structures.

The polarimetric data have shown that the presence 
of equatorial density enhancement is prevalent
among PPNs with widely varying degrees of 
strength.
With the presence of axisymmetry in the late 
AGB mass loss history being established as 
fundamental, it is then reasonable to assume 
that something equally fundamental is responsible 
as the origin for axisymmetry - the equatorial 
enhancement - in the AGB mass loss.
An axisymmetric mass loss process may be a natural 
consequence of very fundamental physics involved in 
any stellar mass loss and/or any dusty outflowing
astronomical phenomena.
In Table \ref{dynamic}, we summarize quantities
of geometric and dynamic nature of the objects.
We measured the inner and superwind (outer) radii
($R_{\rm in}$ and $R_{\rm sw}$, respectively)
assuming the presence of the inner cavity based
on their $P$ structure.
Using the distance and shell expansion velocities 
for these objects taken from the literature
\citep{loup90,omont93,meixner97,reddy99,hony03,hrivnak03,vds03,meixner04},
we derived the duration of superwind ($T_{\rm sw}$) and 
expansion the cessation of mass loss ($t_{\rm dyn}$).
These quantities would not only characterize 
the axisymmetric nature of PPNs but 
also serve as constraining parameters for any 
models to generate equatorial enhancements in 
the AGB mass loss with magnetic fields/companion
objects
(e.g., \citealt{soker98,mm99,matt00}).

\section{Conclusions}

We have performed imaging polarimetry using 
{\hst}/NICMOS to observe optically thin PPNs, 
with the aim to reveal their density structure 
by the polarized intensity, {\ipol}.
In the optically thin regime, the ({\ipol}, $P$\/)
data set can effectively probe the presence of 
scattering bodies distributed in the vicinity 
of the bright central illumination source.
This is a great advantage of imaging polarimetry 
over conventional imaging, in which the central 
star (and its PSF structure) will always dominate 
in these objects and the detection of faint 
nebulosities will not be trivial.

We have found that
(1)
{\iras} 07134 is an equatorially enhanced, 
prolate hollow spheroidal shell that is 
nearly edge-on. 
The presence of the inner cavity has been 
observationally confirmed and the peak 
imbalance seen in the present study and past
study of mid-IR dust emission has been 
determined to be due to density effects,
(2) {\iras} 06530 is a prolate spheroid with
a possible cavity, but its equatorial enhancement 
is more bipolar-like, filling the cavity with 
some matter,
(3) {\iras} 04296 is a combination of two 
spheroids intersecting with each other with 
one of the extensions being inclined towards us,
(4) {\iras} 02229 resembles {\iras} 06530, but 
has greater equatorial enhancement and the 
inclination angle closer to pole-on compared 
with other objects, and
(5) {\iras} 16594 has very bipolar-like morphological
characteristics indicating relatively high optical
depth of the shell, whose multi-polar structure is 
due to a pair of outflows channeled out from
a rotating/precessing torus.
These observations have also indicated that there 
are multiple channels of evolution for multi-polar
shells:
while {\iras} 04296 seems to be a co-existing 
quadrupolar nebula, multiple poles of {\iras} 
16594 likely result from a rotating/precessing 
torus.

Our observations have strongly suggested that PPNs do
possess an inner cavity which physically separates
the shell from the central star and represents 
dynamical time for the
shell expansion since the end of the AGB phase..
The inner PPN shell seems to be density bounded and its
structure is likely shaped by superwind.
In the case of {\iras} 07134, we have been able
to observationally confirm that the inner PPN 
structure is a hollow spheroid with some
equatorial enhancement.
We have also confirmed that SOLE-toroidal PPNs 
have optically thin shells and that the varying 
degree of equatorial density enhancement (i.e., 
the optical depth) determines the detailed shell 
morphology even among this specific group of 
optically thin PPNs. 

\acknowledgements
This research is based on observations with the NASA/ESA 
Hubble Space Telescope, obtained at the Space Telescope 
Science Institute (STScI), which is operated by the 
Association of Universities for Research in Astronomy, 
Inc.\ under NASA contract No.\ NAS 5-26555.
Authors have been supported by  NASA STI 9377.05-A.
Ueta has also been supported by the project IAP P5/36 
financed by FSTC of the Belgian State.
The assistance from E.\ Roye and A.\ Schultz during
data reduction was appreciated.
We thank K.\ Volk who generously provided us with 
mid-IR data taken at Gemini.
F.\ Courbin is also thanked for his help using the MCS 
deconvolution software.
We also appreciate valuable comments from anonymous
referee that helped improving the paper.


\begin{thebibliography}{dummy}

\bibitem[Balick \& Frank(2002)]{balick02}
  Balick, B., \& Frank, A.
  2002, \araa, 40, 439

\bibitem[Bl\"{o}cker(1995)]{bloecker95}
  Bl\"{o}cker, T.
  1995, \aap, 297, 727

\bibitem[Dayal et al.(1998)]{dayal98}
  Dayal, A., Hoffmann, W.\ F., Bieging, J.\ H., Hora, J.\ L., 
  Deutsch, L.\ K., \& Fazio, G.\ G. 
  1998, \apj, 492, 603

\bibitem[Dickinson et al.(2002)]{nicmoshandbook}
  Dickinson, M., et al.\
  2002, in ``HST NICMOS Data Handbook Version 5.0'', ed.\ B.\ Mobasher
  (Baltimore: STScI) 

\bibitem[Garc\'{\i}a-Hern\'{a}ndez et al.(2004)]{gh04}
  Garc\'{\i}a-Hern\'{a}ndez D.\ A., Manchado, A., Garc\'{\i}a-Lario, P.,
  Ben\'{\i}tez Ca\~{n}ete, A., Acosta-Pulido, J.\ A., 
  P\'{e}rez Garc\'{\i}a, A.\ M.
  2004, in ASP Conf.\ Ser.\ Vol.\ 313, 
  Asymmetric Planetary Nebulae III,
  eds.\ M.\ Meixner, J.\ Kastner, B.\ Balick, \& N.\ Soker
  (San Francisco: ASP), 367

\bibitem[Gledhill et al.(2001)]{gledhill01}
  Gledhill, T.\ M., Chrysostomou, A., Hough, J.\ H., \& Yates, J.\ A.
  2001, \mnras, 322, 321

\bibitem[Gledhill \& Takami(2001)]{gledhill01b}
  Gledhill, T. M., \& Takami, M.
  2001, \mnras, 328, 266

\bibitem[Gledhill \& Yates(2003)]{gledhill03}
  Gledhill, T.\ M., \& Yates, J.\ A.
  2003, \mnras, 343, 880

\bibitem[Hines, Schmidt, \& Schneider(2000)]{hines00}
  Hines, D.\ C., Schmidt, G.\ D., \& Schneider, G.
  2000, \pasp, 112, 983

\bibitem[Hony et al.(2003)]{hony03}
  Hony, S., Tielens, A.\ G.\ G.\ M., Waters, L.\ B.\ F.\ M., \& 
  de Koter, A.
  2003, \aap, 402, 211

\bibitem[Hrivnak \& Kwok(1999)]{hrivnak99}
  Hrivnak, B.\ J., \& Kwok, S.
  1999, \apj, 513, 869

\bibitem[Hrivnak \& Reddy(2003)]{hrivnak03}
  Hrivnak, B.\ J., \& Reddy, B.\ E.
  2003, \apj, 590, 1049

\bibitem[Hrivnak, Kwok, \& Su(1999)]{hks99}
  Hrivnak, B.\ J., Kwok, S., \& Su, K.\ Y.\ L.\
  1999, \apj, 524, 849

\bibitem[Iben(1981)]{iben81} 
  Iben, I.\ Jr.
  1981, ApJ, 246, 278

\bibitem[Iben\/(1995)]{iben95}
  Iben, I. Jr.
  1995, Physics Report, 250, 2

\bibitem[Iben \& Renzini(1983)]{iben83}
  Iben, I. Jr., \& Renzini, A.
  1983, \araa, 21, 271

\bibitem[Jura, Chen, \& Werner(2000)]{jura00}
  Jura, M., Chen, C., \& Werner, M.\ W.
  2000, \apj, 544, L141

\bibitem[Koekemoer et al.(2002)]{dithering}
  Koekemoer, A.\ M., et al.\
  2002, ``HST Dither Handbook'', Version 2.0 (Baltimore: STScI)

\bibitem[Kwok(1993)]{kwok93}
  Kwok, S.
  1993, \araa, 31, 63.

\bibitem[Kwok, Su, \& Hrivnak(1998)]{kwok98}
  Kwok, S., Su, K.\ Y.\ L., \& Hrivnak, B.\ J.
  1998, \apj, 501, L117

\bibitem[Kwok, Volk, \& Hrivnak(2002)]{kwok02}
  Kwok, S., Volk, K., \& Hrivnak, B.\ J.
  2002, \apj, 573, 720

\bibitem[Loup et al.(1990)]{loup90}
  Loup, C., Forveille, T., Nyman, L.\ \AA, \& Omont, A.
  1990, \aap, 227, L29

\bibitem[Magain, Courbin, \& Sohy(1998)]{mcs}
  Magain, P., Courbin, F.\, and Sohy, S.
  1998, \apj, 494, 472

\bibitem[Malhotra et al.(2002)]{nicmosihb}
  Malhotra, S., et al.\
  2002, ``NICMOS Instrument Handbook'', Version 5.0 (Baltimore: STScI)

\bibitem[Mastrodemos \& Morris(1999)]{mm99}
  Mastrodemos, N. \& Morris, M.
  1999, \apj, 523, 357

\bibitem[Matt et al.(2000)]{matt00}
  Matt, S., Balick, B., Winglee, R., \& Goodson, A.
  2000, \apj, 545, 965

\bibitem[Meixner et al.(2002)]{meixner02}
  Meixner, M., Ueta, Bobrowsky, M., \& Speck, A.\ K.
  2002, \apj, 571, 936

\bibitem[Meixner et al.(1999)]{meixner99}
  Meixner, M., Ueta, T., Dayal, A., Hora, J.\ L., Fazio, G., 
  Hrivnak, B.\ J., Skinner, C.\ J., Hoffmann, W.\ F., \& Deutsch, L.\ K.
  1999, \apjs, 122, 221

\bibitem[Meixner et al.(1997)]{meixner97}
  Meixner, M., Skinner, C.\ J., Graham, J.\ R., Keto, E., 
  Jerrigen, J.\ G., \& Arens, J.\ F.
  1997, \apj, 482, 897

\bibitem[Meixner et al.(2004)]{meixner04}
  Meixner, M., Zalucha, A., Ueta, T., Fong, D., \& Justtanont, K.
  2004, \apj, accepted

\bibitem[Omont et al..(1993)]{omont93}
  Omont, A., Loup, C., Forveille, T., te Lintel Hekkert, P., 
  Habing, H., \& Sivagnanam, P.
  1993, \aap, 267, 515
 
\bibitem[Reddy, Bakker, \& Hrivnak(1999)]{reddy99}
  Reddy, B.\ E., Bakker, E.\ J., \& Hrivnak, B.\ J.
  1999, \apj, 524, 831

\bibitem[Renzini(1981)]{renzini81}
  Renzini, A.
  1981, in Physical Processes in Red Giants,
  eds.\ I.\ Iben Jr.\ \& A.\ Renzini 
  (Dordrecht: Reidel), 431

\bibitem[Sahai(1999)]{sahai99}
  Sahai, R.
  1999, \apj, 524. L125 

\bibitem[Schr\"{o}der, Winters, \& Sedlmayr(1999)]{schroder99}
  Schr\"{o}der,, K.-P., Winters, J.\ M., \& Sedlmayr, E.
  1999, \aap, 349, 898

\bibitem[Soker(1998)]{soker98}
  Soker, N.
  1998, \mnras, 299, 1242

\bibitem[Sparks \& Axon(1999)]{sparks99}
  Sparks, W.\ B., \& Axon, D.\ J.
  1999, \pasp, 111, 1298

\bibitem[Steffen \& Sch\"{o}nberner(2000)]{steffen00}
  Steffen, M., \& Sch\"{o}nberner, D.
  2000, \aap, 357, 180

\bibitem[Steffen, Szczerba, \& Sch\"{o}nberner(1998)]{steffen98}
  Steffen, M., Szczerba, R., \& Sch\"{o}nberner, D.
  1998, \aap, 337, 149

\bibitem[Su, Hrivnak, \& Kwok(2001)]{su01}
  Su, K.\ Y.\ L., Hrivnak, B.\ J., \& Kwok, S.
  2001, \aj, 122, 1525

\bibitem[Su et al.(2003)]{su03}
  Su, K.\ Y.\ L., Hrivnak, B.\ J., Kwok, S., \& Sahai, R.
  2003, \aj, 126, 848

\bibitem[Ueta \&  Meixner(2003)]{ueta03}
  Ueta, T., \& Meixner, M.
  2003, \apj, 586, 1338

\bibitem[Ueta, Meixner, \& Bobrowsky(2000)]{ueta00}
  Ueta, T., Meixner, M., \& Bobrowsky, M.
  2000, \apj, 528, 861

\bibitem[Ueta et al.(2001)]{ueta01}
  Ueta, T., Meixner, M., Hinz, P.\ M., Hoffmann, W.\ F., 
  Brandner, W., Dayal, A., Deutsch, L.\ K., Fazio, G.\ G., \& 
  Hora, J.\ L.
  2001, \apj, 557, 831

\bibitem[Ueta et al.(2003)]{ueta03a}
  Ueta, T., Meixner, M., Moser, D.\ E., Pyzowski, L.\ A.,
  \& Davis, J.\ S. 
  2003, \aj, 125, 2227

\bibitem[Ueta et al.(2004)]{ueta04}
  Ueta, T., de Vlieger, M., Van de Steene, G.\ C., Szczerba, R.,
  \& Van Winckel, H.
  2004, \aap, submitted

\bibitem[Van de Steene \& van Hoof(2003)]{vds03}
  Van de Steene, G.\ C., \& van Hoof, P.\ A.\ M.
  2003, \aap, 406, 773

\bibitem[Van Winckel(2003)]{vanwinckel03} 
  Van Winckel, H.
  2003, \araa, 41, 391 

\bibitem[Whitney \& Hartmann(1993)]{wh93}
  Whitney, B.\ A., \& Hartmenn, L.
  1993, \apj, 402, 605
\end{thebibliography}
\end{document}